\documentclass[useAMS,usenatbib,usegraphicx]{mn2e}
\usepackage{journals}
\usepackage{times}
\usepackage{color}
\title[Total mass profiles of ellipticals]
{Dark matter in
elliptical galaxies:
I. Is the total mass density profile of the NFW form or even steeper?}

\author[Gary A. Mamon
and Ewa L. {\L}okas]{Gary A.
    Mamon$^{1,2}$\thanks{E-mail: gam@iap.fr} and Ewa
    L. {\L}okas$^3$\thanks{E-mail: lokas@camk.edu.pl}\\ $^1$Institut
d'Astrophysique de
    Paris (UMR 7095: CNRS \& Univ. Pierre \& Marie Curie), 98 bis Bd Arago,
F--75014 Paris, France \\ 
$^2$GEPI (UMR 8111: CNRS \& Univ. Denis Diderot), Observatoire de Paris,
    F--92195 Meudon, France\\
$^3$Nicolaus Copernicus Astronomical Center, Bartycka 18,
    PL--00716 Warsaw, Poland\\
}

\date{Accepted ???. Received ????; in original form ????}
\pubyear{2005}

\begin{document}

\maketitle

\begin{abstract}

Elliptical galaxies are modelled as S\'ersic luminosity distributions with
density profiles (DPs) for the total mass adopted from the DPs of haloes
within dissipationless $\Lambda$CDM N-body simulations. Ellipticals 
turn out to be
inconsistent with cuspy low-concentration NFW models representing the total
mass, nor are they consistent with a steeper $-1.5$ inner slope, nor with the
shallower models proposed by Navarro et al. (2004), nor with NFW models 10 times
more concentrated than predicted, as deduced from several X-ray observations:
the mass models, extrapolated inwards, lead to local mass-to-light ratios 
that are smaller
than the stellar value inside an effective radius ($R_e$), and to central
aperture velocity dispersions that are much smaller than observed. This
conclusion remains true as long as there is no sharp steepening (slope $< -2$)
of the dark matter (DM) DPs just inside 0.01 virial radii. 

The too low total mass and velocity dispersion produced within $R_e$ by 
an NFW-like total mass profile
suggests that 
the stellar component should 
dominate the DM one out to at least $R_e$. It should then be
difficult to 
kinematically constrain the inner slope of the dark matter DP of
ellipticals. The
high concentration parameters deduced from X-ray observations appear to
be a consequence of fitting an NFW model to the total mass DP made up of a
stellar component that dominates inside and a DM component that dominates
outwards. 

An appendix gives
the virial mass dependence of the concentration parameter, 
central density, and total mass of the
Navarro et al. model. In a 2nd appendix are given single integral expressions
for the velocity dispersions averaged along the line-of-sight, in circular
apertures and in thin slits, for general luminosity density and mass
distributions, with isotropic orbits.

\end{abstract}


\section{Introduction}

There has been much recent progress on the photometric characterization of
elliptical galaxies. Whereas a variety of models for the optical surface
brightness  profiles of ellipticals have been used in the past, such as the
Hubble-Reynolds law \citep{Reynolds1913},
the analytical \cite{King62_clusters} or modified Hubble law,
the projection of the isothermal spheres truncated in phase space 
\citep{King66}
and the $R^{1/4}$ law \citep{deVaucouleurs48}, none provides
an adequate fit to the surface photometry
of the large majority of elliptical galaxies.
However, there has
been a recent consensus on the applicability to virtually all elliptical
galaxies \citep*{CCDO93,BCDP02}
of the generalization (hereafter, S\'ersic law) of the $R^{1/4}$ law proposed
by \cite{Sersic68}, which can be written as
\begin{equation}
I(R) = I_0 \exp \left [-\left ({R \over a_S}\right )^{1/m} \right ] \ ,
\label{Isersic}
\end{equation}
where $I$ is the surface brightness, $a_S$ the S\'ersic scale parameter, and
$m$ the S\'ersic shape parameter, with $m=4$ recovering the $R^{1/4}$ law.
Moreover, strong correlations have been reported between the shape parameter
$m$ and either luminosity or \emph{effective} (half projected light)
radius $R_e$ (\citealp{CCDO93}, see also \citealp{PS97}, and references
therein).

On the other hand, there is
still much uncertainty on the importance of dark matter in elliptical
galaxies, especially in their outer regions.
Even in the spherical case considered in this paper,
kinematical modelling of ellipticals
usually cannot disentangle the degeneracy between
the uncertainty of their
gravitational potential and that of their internal kinematics (the anisotropy
of their velocity ellipsoid), unless velocity profiles or at least 4th order
velocity moments are considered 
\citep{Merritt87,RW92,vdMF93,Gerhard93,LM03,KBM04}.

 
Analyses of diffuse X-ray emission in elliptical galaxies have the advantage
that the equation of hydrostatic equilibrium has no anisotropy term within
it, so in spherical symmetry, one easily derives the total mass distribution
through (e.g. \citealp{FLG80})
\begin{equation}
M(r) = - {k\,T(r)\, r \over G \mu m_p}\, 
\left ({{\rm d}\ln n\over {\rm d}\ln r} + {{\rm d}\ln T \over {\rm d}\ln r}
\right ) 
\ ,
\label{MofrX}
\end{equation}
where $k$ is the Boltzmann constant, while $T$, $n$ and $\mu
m_p$ are respectively 
the temperature, electron density and mean particle mass of the
plasma.
However, it is crucial to measure $T(r)$ and its gradient, and 
unfortunately, even with the two new generation X-ray telescopes
{\it XMM-Newton} and 
{\it Chandra}, it is difficult to achieve such measurements beyond some
fraction of 1/2 the virial radius for galaxy clusters \citep{Arnaud+02,PA02}
but much less for elliptical galaxies.
Moreover, the X-ray emission from elliptical galaxies is the combination of
two components: diffuse hot gas swimming in the gravitational potential as
well 
as direct emission from individual stars, and it can be highly difficult to
disentangle the two (see \citealp{BB01}).


On the theoretical front,
cosmological simulations of large chunks of the Universe dominated by cold
dark matter (CDM) have
recently
reached enough mass and spatial resolution that there appears to be a
convergence on the structure and internal kinematics of the bound
structures,
usually referred to as \emph{haloes}, in the simulations.
In particular, the density profiles appear to converge
to one with an outer slope of $\simeq -3$ and an inner slope between $-1$ 
(\citealp*{NFW95,NFW96})
and $-3/2$ \citep{FM97,Moore+99,Ghigna+00}, see also
\cite{Power+03,FKM04}.

In this paper, we consider 
the general formula that \cite{JS00} found to provide
a good fit to simulated haloes:
\begin{equation}
\rho(r) \propto \left ({r\over a_h} \right)^{-\alpha} \,\left [1 + \left
({r\over a_h}\right )\right ]^{\alpha-3} \ ,
\label{rhodark}
\end{equation} 
where $\alpha = 1$ (hereafter `NFW')
or
3/2 (hereafter `JS--1.5'),
and `$h$' stands for
\emph{halo}.
These profiles fit well the density profiles of cosmological simulations out
to the \emph{virial} radius $r_v$, wherein the mean density is
$\Delta \approx 200$ times the critical density of the Universe.\footnote{For
the standard $\Lambda$CDM parameters, 
$\Omega_m=0.3,\Omega_\lambda=0.7$, the approximate formulae of 
\cite{KS96_ApJ} and \cite{BN98} yield $\Delta\simeq100$ (see also \citealp{ECF96} and
\citealp{LH01}), where 
the exact value is 101.9.
However, many cosmologists prefer 
to work with the value of 200, which is close to the value
of 178 originally derived for the Einstein de Sitter Universe ($\Omega_m=1,
\Omega_\lambda=0$).} 
The ratio of virial radius to scale radius is called the \emph{concentration
  parameter}:
\begin{equation}
c={r_v \over a_h} \ .
\label{cdef}
\end{equation}

Very recently, a number of studies have proposed better analytic fits to the
density
\citep{Navarro+04,DMS04_rho} or circular velocity \citep{SWTS02,Stoehr05}
profiles of simulated haloes.
In particular, the formula of 
\citeauthor{Navarro+04} 
\begin{equation}
{d \ln\rho\over d\ln r} = -2\,\left ({r\over r_{-2}} \right)^{1/\mu}
\ ,
\label{Nav04slope}
\end{equation}
where $\mu \simeq 6$ and $r_{-2}$ is the radius of slope $-2$,
is attractive because it
converges to a finite central density at very small scales and 
steepens progressively at outer radii to produce 
a finite mass. Moreover, \citeauthor{Navarro+04}
obtained their fits
to the logarithmic slope of the density profile, while \citeauthor{DMS04_rho}
fit to the density profile and Stoehr and collaborators to the circular
velocity profile. 
Since the density and circular velocity respectively
involve single and double integrals
of the logarithmic slope profile, the latter two can mask subtle variations
only picked up in the logarithmic slope.
For these reasons, we shall add the \citeauthor{Navarro+04} model in the
present work.


In a previous paper 
(\citealp{LM01}, hereafter paper I), 
we showed
that projected  NFW density profiles could, in principle, be
fit to S\'ersic profiles with $m\simeq 3$.
However, there are three reasons to disregard this match as a proof that
mass
follows light throughout elliptical galaxies:
\begin{enumerate}
\item \label{mfitNFW}
As shown in Fig. 12 of paper I,
projected NFW density profiles fit the S\'ersic profiles only within
a narrow range of S\'ersic shape parameters $2.7 \leq m \leq 4.0$, given
reasonable NFW
concentration parameters 
with ($5 \leq c \leq 22$), whereas
ellipticals are fit by S\'ersic profiles in the much wider range:
from $m = 0.5 - 0.6$ for cluster dwarf ellipticals 
\citep{CCDO93,BJ98,Marquez+00}
to $m = 16$ (\citeauthor{CCDO93})
or 8 \citep{Graham98}, 7 \citep{DOnofrio01}
or at least 5.6 \citep{Marquez+00}.
\item \label{normfitNFW}
The fits produced enormous effective radii, much larger than observed.
\item 
\label{parcorfit}
The fits produced $m$ increasing with concentration parameter $c$
(again Fig.~12 of paper~I --- which plots $1/m$ versus $c$).
But, $c$ is known to decrease with mass within the virial radius 
\citep*{NFW97,JS00}, while $m$
is known to increase with luminosity (\citeauthor{CCDO93}; \citealp{PS97},
see Fig.~\ref{figcor} below). 
Hence, one arrives at the very unlikely result
that galaxy luminosity decreases with increasing mass within the virial
radius.
\end{enumerate}
The conclusions of paper~I have been confirmed in a recent analysis of
\cite{MNLJ05}, who have recently shown that the halos in
dissipationless 
cosmological $N$-body simulations are also well fit in projection by a
S\'ersic model with $m\simeq 3$. 
This is not surprising since the NFW model provides an adequate fit to the
density profiles of halos in dissipationless cosmological simulations.
But, in contrast to the case of the divergent-mass NFW models, 
point~\ref{normfitNFW} should not be relevant for convergent-mass halos
described by the S\'ersic or 
\citeauthor{Navarro+04} models. Nevertheless,
the relevance of points~\ref{mfitNFW} and \ref{parcorfit}
remain relative to the work of \citeauthor{MNLJ05}.

Now there are several strong indications that mass does not follow
light in elliptical galaxies: first, from the kinematics of neutral gas
\citep{BPPS93}, and second from X-rays (eq.~[\ref{MofrX}]), albeit
with the simplifying and probably optimistic assumption of isothermality.
Some studies point to $M/L_{\rm opt}$ increasing outwards,
such as \cite{Jones+97} for NGC 1399.
Also, \cite{BJCG02} use the variation of the ellipticity of the
X-ray isophotes to conclude 
that mass does not follow light in an elongated elliptical galaxy
(NGC 720).

Recent analyses of X-ray data by \cite{Sato+00}, \cite{WX00}, and
\cite{LP02}
(quoted in \citealp{LBP02}) conclude (using
eq.~[\ref{MofrX}]) that the 
total mass of ellipticals, groups and clusters are indeed well fit by an NFW
model, but with typically ten times larger concentration
than measured 
in cosmological simulations for the
mass range of elliptical galaxies.
Surprisingly, there have been no detailed studies of these high concentration
total density profiles arising from the X-ray observations.

Whereas the NFW profile was first established with cosmological simulations
including a dissipative gaseous component \citep{NFW95}, most confirmations
of the NFW or JS--1.5 profiles have come from simulations without gas,
which can achieve a much greater mass resolution.  
Hence, one may ask if the density profiles of haloes
found in cosmological $N$-body
simulations of cold dark matter in a flat universe with $\Omega_\Lambda
\simeq 0.7$ (globally referred hereafter as $\Lambda$CDM)
represent the \emph{total} density of observed cosmic structures
or only their dark matter component.
Given that the average 
baryon fraction in the Universe is small, one could think that when
structures such as elliptical galaxies form by collapse, the baryons simply
follow the 
dark matter without affecting it, and the total mass profile would resemble
the low concentration cuspy models found in the cosmological simulations.
Alternatively, since gas dynamics is dissipative, there may be a density
threshold beyond which the gas will decouple from the dark matter, and then
the central regions of ellipticals would be dominated by gas and later stars
that form from it. The low-concentration cuspy density profiles found in
the cosmological simulations would then only apply to the dark matter.
Moreover, if the baryons dominate the dark matter at small radii, the dark
matter will re-adjust itself within the gravitational potential dominated by
the baryons (in a process often referred to as adiabatic contraction, see,
e.g., \citealp{GKKN04}), so that even the dark matter density profile may differ
substantially from the predictions of the dissipationless $\Lambda$CDM
simulations.
Also, if ellipticals form by major mergers of spirals
\citep{Toomre77,M92}, the distribution of 
baryons (mainly stars) will be set by violent relaxation operating during the
merger, and if the baryonic fraction is low in the outer regions of the
two merging spirals, the merger remnant should also show a lower baryon
fraction in the outer regions.

In the present paper, we
abandon the hypothesis of paper~I of constant
mass-to-light ratio, and we consider the next simplest approach, asking 
ourselves whether the radial profiles of 
density
coming out of dissipationless
cosmological $N$-body simulations are consistent with the observations 
(surface photometry and spectroscopy) of elliptical galaxies,
for the \emph{total} mass (Sec.~\ref{results}),
and whether low or high
concentration parameters are required.
We will not consider here the response of the dark matter component to the
dissipative baryonic component (e.g., the adiabatic contraction of the dark
matter).

We begin, in Sec.~\ref{basic}, with a summary of the luminosity
and mass models of elliptical galaxies that we adopt in this paper.
In a companion paper \citep{ML05b},
we go one step further and confront the general observed trends in 
elliptical galaxies with the predictions from a 4-component model of
ellipticals with stars, dark matter, hot gas and a central black hole,
allowing for slight radial velocity anisotropy, as seen in cosmological
$N$-body simulations.

\section{Basic equations}
\label{basic}


\subsection{Distribution of optical light}
\label{dislight}

The S\'ersic (eq.~[\ref{Isersic}]) optical surface brightness profile that
represents the projected stellar distribution, can be deprojected
according to the
approximation first proposed by \cite{PS97}
\begin{eqnarray}
\ell(r) &=& \ell_1\,\widetilde \ell(r/a_S) \label{nuofr}\\
\widetilde \ell(x) &\simeq&
x^{-p}\,\exp \left (-x^{1/m}\right ) \ , \label{nuofx} \\
\ell_1 &=& \left \{{\Gamma(2m)\over \Gamma[(3-p)\,m]}\right\}\,{I_0 \over
2\,a_S} \ , \label{nu1} \\
p &\simeq& 1.0 - 0.6097/m + 0.05463/m^2 \ , \label{pofm}
\end{eqnarray}
where $\Gamma(a)$ is the gamma function.
The last equation is from 
\cite{LGM99}
who argue that
equations~(\ref{nuofr}) and (\ref{nuofx}) then 
provide a better deprojection of
the S\'ersic profile (eq.~[\ref{Isersic}]),
and in a wider range of $m$ (good to better than 5\% accuracy for $0.55 < m <
10$ 
within $0.01 < R/R_e < 1000$, 
where
over 99.5\% of the light lies)
than a previous approximation of $p(m)$
proposed by \citeauthor{PS97}.

The integrated luminosity corresponding to
equations~(\ref{nuofr}), (\ref{nuofx}) and (\ref{nu1}) is then
(\citeauthor{LGM99})
\begin{eqnarray}
L_3 (r) &=& L\, \widetilde L_3(r/a_S) \label{L3ofr} \\
\widetilde L_3(x) &=& {\gamma \left [(3-p)m, x^{1/m}\right ]\over
\Gamma[(3-p)m]} \label{L3tilde} \ ,
\end{eqnarray}
where $\gamma(a,x)$ is the incomplete gamma function and
where the total luminosity of the galaxy is 
\begin{equation}
L = 2\pi m\,\Gamma(2m)\,I_0\, a_S^2 = 4\pi m\,
\Gamma\left [(3-p)\,m\right ]\,\ell_1\,a_S^3\ ,
\label{Ltot}
\end{equation}
as obtained by \cite{YC94} from the S\'ersic surface brightness
profile of  
equation~(\ref{Isersic}), and which matches exactly the total
luminosity obtained  by integration of \citeauthor{LGM99}'s
approximate deprojected profile.

\begin{figure*}
\includegraphics[width=0.9\hsize]{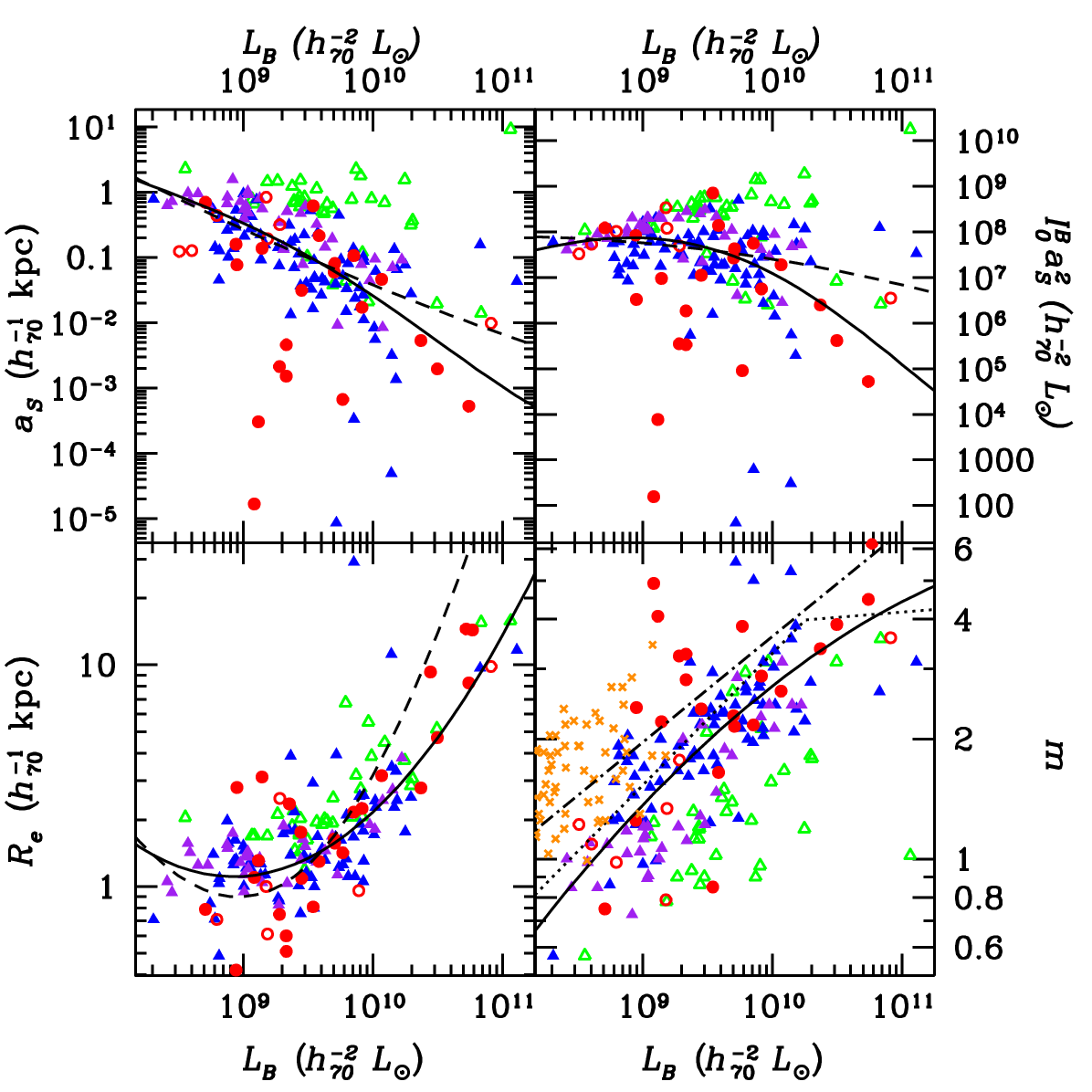}
\caption{Correlations of S\'ersic parameters of elliptical and S0 galaxies.
Ellipticals in the Coma and Abell 496 clusters (\emph{filled triangles}) and
Abell 85 (\emph{empty triangles})
measured by 
Marquez et al. (2000);
ellipticals (\emph{filled circles}) and S0s
(\emph{open circles})
in the Virgo and Fornax clusters  
measured by 
D'Onofrio (2001, restricted to the good-fitting 
cases, i.e. where $Q<3$ and $\chi_r^2<1.5$)
and dwarf ellipticals in the Virgo cluster  (\emph{crosses})
and in the $m$ vs. $L$ plot 
measured by 
Binggeli \& Jerjen (1998).
These symbols are plotted in \emph{blue}, \emph{purple}, \emph{green},
\emph{red}, and \emph{orange} in the electronic version of the Journal. 
The \emph{solid curves} are the adopted fits,
performed on $\log R_e$ vs. $\log L$ (eq.~[\ref{Refit}])
and $\log m$ vs. $\log L$ (eq.~[\ref{mfit}]), 
and used to estimate $a_S$ (eq.~[\ref{aSfit}]) and $I_0\,a_S^2$
(eq.~[\ref{Ltot}]), while 
the \emph{dashed curves} represent the direct 
fits of $\log a_S$ vs. $\log L$ and $\log I_0\,a_S^2$
vs. $\log L$ and the corresponding relation for $R_e$ vs. $L$ (from
eqs.~[\ref{Re}] and [\ref{bofm}]).
The fits use the data of \citeauthor{Marquez+00} omitting their Abell~85
values, combined with the data of \citeauthor{DOnofrio01}.
The
\emph{dotted broken line} in the $m$ vs. $L$ plot 
is the mean relation from 
Prugniel \& Simien (1997), while the \emph{dashed-dotted line} is from
Graham \& Guzman (2003).
The luminosities of the D'Onofrio galaxies are $H_0$-independent (see text).
}
\label{figcor}
\end{figure*}
\nocite{GG03}
Fig.~\ref{figcor} shows how the S\'ersic parameters are correlated.
The data are taken from 
\begin{enumerate}
\item
\cite{BJ98}, who computed S\'ersic parameters from fitting the
cumulative luminosity profile for dwarf ellipticals in Virgo.
\item 
\cite{Marquez+00}, who did the same for normal ellipticals 
in the Coma, Abell 85 and Abell
496 clusters.
We converted their angular sizes to
distances, by simply assigning Hubble distances to each cluster, 
assuming no peculiar
motion
relative to us. 
We also converted their $V$-magnitudes to $B$-band luminosities, assuming
$B-V = 0.96$ (typical of elliptical galaxies, e.g. \citealp{FSI95})
and (e.g. \citealp{CBC96})
\begin{equation}
M_B^\odot = 5.45 \ .
\label{MBsun}
\end{equation}
\item
\cite{DOnofrio01}, who performed 2D fits of S\'ersic + exponential models for
galaxies in the Virgo and Fornax clusters.
We corrected their luminosities by replacing their 
unique distance modulus  by
the surface brightness fluctuation (SBF) distance modulus 
of \cite{Tonry+01} when available, or otherwise by adding the
median 
difference (0.18) between the SBF distance moduli and the unique value
adopted by \citeauthor{DOnofrio01} for their 23 galaxies in common.
\end{enumerate}

In principle, the 2D fitting 
method of \citeauthor{DOnofrio01} should be the most
reliable, but for given clusters, the means trends in the data of
\citeauthor{Marquez+00}  show
much less scatter for
individual clusters.
Interestingly, the mean trends for
Coma and Abell 496 agree well with 
the mean trend for the ellipticals and lenticulars of
\citeauthor{DOnofrio01}, while 
the galaxies in the cluster Abell 85 (\emph{open 
triangles} in
Fig.~\ref{figcor}) are systematically offset from the galaxies of the other
two clusters, in such a manner that its
distance, as simply derived from its redshift, appears 80\% too
large (as is clear from the correlations of $m$ vs. $L$, 
and consistent with the correlations in the other plots of
Fig.~\ref{figcor}), but it is highly improbable that the distance to Abell 85
is overestimated by such a large factor.
Also, the dwarf ellipticals analysed by \cite{BJ98} appear to have a
shape-luminosity relation that disagrees with those of 
\citeauthor{Marquez+00} and \citeauthor{DOnofrio01}.

Discarding the Abell~85 data as well as the dwarf ellipticals of \cite{BJ98},
one is left (\emph{circles} and \emph{filled triangles}) 
with fairly strong correlations in Fig.~\ref{figcor}, 
where, in particular, the shape
parameter $m$ correlates with luminosity \citep{CCDO93}.

For simplicity, \emph{we assume that 
elliptical galaxies 
constitute a one-parameter family, based on luminosity} (or equivalently, on
S\'ersic shape 
$m$).
Whereas such a one-parameter model of ellipticals is consistent with the
Faber-Jackson (\citeyear{FJ76}) relation between central
velocity dispersion and luminosity, it
is obviously an oversimplification, in view of the fundamental plane 
of elliptical galaxies, where the central velocity dispersion is a function
of both luminosity \emph{and} surface brightness \citep{Dressler+87,DD87}.

We 
fit the parameter correlations of Fig.~\ref{figcor} 
with 2nd-order polynomials in log space (with iterative 3-$\sigma$ rejection
of outliers), using the 
galaxies in Coma and Abell 496 of \citeauthor{Marquez+00} (\emph{filled
triangles} in Fig.~\ref{figcor}) and the ellipticals and S0s in Virgo of
\citeauthor{DOnofrio01} (\emph{circles} in Fig.~\ref{figcor}).
Now $a_S$ and $R_e$ are directly related through $m$ via
\begin{eqnarray}
R_e &=& b^m\,a_S \label{Re}\\
b &\simeq& 2\,m - 1/3 + 0.009876/m \ , \label{bofm}
\end{eqnarray}
where the latter relation is from \cite{PS97}.
Similarly,
$L$ and $I_0\,a_S^2$ are directly related through $m$ (eq.~[\ref{Ltot}]).
Therefore, we choose to obtain $a_S$ and $I_0\,a_S^2$ through the fits of 
$R_e$ vs. $L$, and $m$ vs. $L$:
\begin{eqnarray}
\log h_{70}\,R_e^{\rm (fit)} &\!\!\!\!=\!\!\!\!& 
0.34 + 0.54\,\log L_{10} + 0.25\,\left (\log
L_{10}\right)^2 \ , 
\label{Refit}\\
\log m^{\rm (fit)} &\!\!\!\!=\!\!\!\!&
0.43 + 0.26\,\log L_{10} - 0.044\,\left (\log L_{10} \right)^2 \,,
\label{mfit}
\end{eqnarray}
where $L_{10} = h_{70}^2\,L_B / \left ( 10^{10} {\rm L_\odot} \right)$, 
$R_e^{\rm (fit)}$ is measured
in kpc, and
with $H_0 = 100\,h = 70\,h_{70} \, \rm km \,
s^{-1}
\,Mpc^{-1}$.
Then equations~(\ref{Re}) and (\ref{bofm}) lead to
\begin{equation}
a_S \simeq \left [b \left(m^{\rm
(fit)}\right)\right]^{-m^{\rm (fit)}} R_e^{\rm (fit)} \ .
\label{aSfit}
\end{equation}

The fits of equations~(\ref{Refit}), (\ref{mfit}) and the corresponding
relation for $a_S$ from equation (\ref{aSfit}) are shown as \emph{solid curves}
in Fig.~\ref{figcor}, whereas the \emph{dashed curves} show the direct fits
to $L$ for $a_S$ and $I_0\,a_S^2$.
For example, for $L_B = 10^{10} h_{70}^{-2} {\rm L_\odot}$, one has $m^{\rm (fit)}
= 2.7$, slightly lower than the value of $m=3.29$ inferred from the relation
of \cite{PS97}.
The fits can be considered to be uncertain to a factor 2 in $R_e$ and 1.5 in
$m$ 
and we will propagate these uncertainties later in our
analysis.
Note that equation~(\ref{Refit}) produces effective radii that increase at
decreasing luminosity for low $L$, which may not be very realistic.
Similarly, equation~(\ref{mfit}) produces abnormally low values of $m$ at low
luminosity, as the $m$ vs $L$ relation should flatten out at low luminosities
to accommodate the dwarfs of \citeauthor{BJ98} and because values of $m <
1$ are rarely mentioned in the literature if at all.
However, inspection of Fig.~\ref{figcor} indicates that
equations~(\ref{Refit}) and (\ref{mfit}) are both reasonably accurate for 
$L_B > 2 \times 10^9 h_{70}^{-2} {\rm L_\odot}$.


\subsection{Scalings of global properties}
\label{scalings}
We adopt a fiducial luminosity of $L_B = L_B^*$ (the luminosity at the break
of the field galaxy luminosity function).
\cite{Liske+03} have compiled various measurements of the corresponding
absolute magnitude $M^*$ and converted to
their $B_{\rm MGC}$ band. Their median value is $M_{\rm MGC}^* = -19.6 + 5\,\log\,h =
-20.37$ (for $h = 0.7$), which is 
within 0.01 magnitude of their conversion of the {\it
2dFGRS} value of \cite{Norberg+02}. 
\citeauthor{Liske+03} give the transformation from
the $B_{\rm MGC}$ system to the Landolt $B$ system: $B_{\rm MGC} = B -
0.145\,(B-V)$.\footnote{The issue of what is the standard $B$ band is
confusing, as \citeauthor{Liske+03}'s conversion from $B_{\rm MGC}$ to
$B_{\rm RC3}$ is quite different from 
their conversion to the Landolt $B$ system.}
Given that elliptical galaxies have $B-V \simeq 0.96$ \citep{FSI95}, 
we derive 
$M_B^* = -20.24$, which corresponds to $L_B^* = 1.88 \times 10^{10}\rm
L_\odot$, 
again using $M_B^\odot = 5.45$ (eq.~[\ref{MBsun}]).

Given the mean luminosity density of the Universe $j$, and the mean mass
density of the Universe $\Omega_m\,\rho_{\rm crit}$, 
the mean mass-to-light ratio 
of the Universe is 
\begin{equation}
\overline \Upsilon = \Omega_m\,\Upsilon_{\rm closure} =
{3\,\Omega_m H_0^2 \over 8\pi\,G\,j} \ .
\label{defmluniv}
\end{equation}
Converting \citeauthor{Liske+03}'s $j_{b_J} = 1.99 \times 10^8 h\,\rm L_\odot
\,Mpc^{-3}$ to the $B$ band with 
$B-b_J=0.28\,(B-V)$, initially proposed by \cite{BG82} and 
inferred from both \citeauthor{Liske+03} and
\cite{Blanton+03}, now assuming $\langle B-V \rangle = 0.94$
\citep{Norberg+02}, with $M_{b_J}^\odot = 5.3$ (\citeauthor{Liske+03}), and
converting the Hubble constant,
we derive $j_B / (10^8\,h_{70}\,{\rm L_\odot}\,\rm Mpc^{-3}) =
1.25$. 
Similarly, \cite{Blanton+03} give blue band luminosity densities of 
$-14.98 - 2.5\,\log h$
and $-15.17 - 2.5\,\log h$
magnitude per Mpc$^3$, when k-corrected to $z=0$ and $z=0.1$, respectively.
For the local Universe ($z=0$),
this translates to $j_B = 1.04 \times 
10^8\,h_{70}\,{\rm L_\odot}\,\rm Mpc^{-3}$,
again using equation~(\ref{MBsun}).
Since \citeauthor{Liske+03} do not k-correct to $z=0$ and since the median
redshift of the {\it 2dFGRS} is close to 0.1, we adopt a mean k-correction
similar 
to the one found by \citeauthor{Blanton+03}: $-14.98 + 15.17 = 0.19$ mag.
This then leads to a {\it 2dFGRS} luminosity density, transformed to the $B$
band 
and k-corrected to $z=0$ of 
$j_B = 1.05 \times 10^8\,h_{70}\,{\rm L_\odot}\,\rm Mpc^{-3}$, in remarkable
agreement with the SDSS value of \citeauthor{Blanton+03}.
Taking the mean of these {\it 2dFGRS} and {\it SDSS} luminosity densities, we
derive 
a local closure mass-to-light ratio of $\Upsilon_{\rm closure}^B/h_{70} =
1300$ and, with $\Omega_m = 0.3$, we obtain
\begin{equation}
\overline \Upsilon_B = 390\,h_{70} \ .
\label{mluniv}
\end{equation}
%
So for $L_B = L_B^* = 1.88 \times
10^{10}\,h_{70}^{-2}\,{\rm L_\odot}$, an unbiased universe would yield
a total mass within the virial radius
$\log h_{70}\,M_t = 12.87$.

Two recent statistical analyses of galaxy properties
 suggest that $M/L_B$ has a non-monotonous
variation with mass (or luminosity), with a minimum value around 100 for
luminosities around $10^{10}\,{\rm L_\odot}$ \citep{MH02,YMvdB03}.
Theoretical predictions by \cite{KCDW99}, 
based upon semi-analytical modelling of galaxy
 formation on top of 
cosmological $N$-body simulations also find a minimum of
$\Upsilon_B / \overline \Upsilon_B = 0.19$ at a similar luminosity, 
which translates to $\Upsilon_B = 74\,h_{70}$ with equation~(\ref{mluniv}).
On the other hand, the internal kinematics of galaxy clusters are consistent
with the mass-to-light ratio of the Universe given by equation~(\ref{mluniv}):
e.g. \cite{LM03} derive
$\Upsilon_B=351\,h_{70}$ for the Coma cluster. 

In the following section, we will compare 
our mass-to-light ratios to those expected for old stellar
populations.
For example, \cite*{TBB04} report a stellar mass-to-blue-light ratio
$\Upsilon_{*,B} = 7.1$ from an analysis of 
SDSS 
galaxies using the {\it PEGASE} \citep{FRV97} stellar population synthesis
code and a \cite{Salpeter55} initial mass function (IMF).
However, using different stellar population synthesis codes and several
choices for the IMF,
\cite{GKSB01} find values differing by a factor of 2 for
given galaxies (e.g., $\Upsilon_{*,B} = 4.2$ to 8.8 for the nearby
elliptical, NGC~3379).
We performed our own tests of the expected values of $\Upsilon_{*,B}$ given
the extinction-corrected colours of nearby elliptical galaxies, with {\it
PEGASE}, {\it GALEXV} \citep{BC03} and G. Worthey's on-line model\footnote{{\tt
http://astro.wsu.edu/worthey/dial/dial\_a\_model.html}, see
\citealp{Worthey94}}. 
We found a very large variety of stellar mass-to-light ratios for given
colours, 
spanning from $\Upsilon_{*,B} = 2$ to 12, depending on the IMF, its low-
and high-mass 
cutoffs, the metallicity and the stellar evolution code.
In particular, \citeauthor{Salpeter55}'s IMF produces high values (as also
found by \citeauthor{GKSB01}), and {\it
GALEXV} produces lower values for a given IMF in comparison with {\it
PEGASE}. In the end, we adopt best values of $\Upsilon_{*,B} = 8$ and 5, and
a possible range of a factor 1.5 around the geometric mean ($\Upsilon_{*,B} =
6.3$).

We also make use of the Faber-Jackson (\citeyear{FJ76}) relation.
We have compared the calibrations of \cite{dVO82}, 
\cite{BZB96},
\citeauthor{Kormendy+97} (\citeyear{Kormendy+97}, who make use of data from
\citealp{McElroy95}) and \citeauthor{ForbesP99} (\citeyear{ForbesP99}, who
make use of data from  
\citealp{PS96}).
Converting to the same value of $H_0$, we find that 
the correlation between the velocity dispersions averaged in circular
apertures, $\sigma_{\rm ap}$ and the blue-band luminosities of
\citeauthor{BZB96} are consistent with  
the analogous correlations of \citeauthor{dVO82},
\citeauthor{Kormendy+97} and \citeauthor{ForbesP99}, but with less scatter,
and extends to a large enough range of luminosities,
with the following mean trend
\begin{equation}
\sigma_{\rm ap} = 
171 \, {\rm km \, s}^{-1} \,\left ({h_{70}^2 \,L_B \over 10^{10}
\,{\rm L_\odot}} \right )^{1/3.25} 
\ .
\label{fjbender}
\end{equation}

\subsection{Distribution of total mass}

We consider here 3  models for the total mass distribution: 
1) the NFW model with inner slope $-1$;
2) 
a generalization of
the NFW model introduced by \cite{JS00} with inner slope $-3/2$
(JS--1.5, which is slightly different from the formula proposed by
\citealp{Moore+99});  
and 3) the convergent model of \citeauthor{Navarro+04}
(\citeyear{Navarro+04}, hereafter Nav04, see eq.~[\ref{Nav04slope}]).
The total matter density profile can generally be written:
\begin{eqnarray}
\rho(r) &\!\!\!\!=\!\!\!\!& {c^3\over g(c)}\,
\left ({M_v\over 4\pi\,r_v^3}\right)\,
\widetilde \rho(r/a_h) \
,
\label{rhodark2}
\\
\widetilde \rho(y) &\!\!\!\!=\!\!\!\!& 
\left \{
\!\!
\begin{array}{lll}
y^{-\alpha}\,(1+y)^{\alpha-3} 
& \hbox{(NFW, JS--1.5)} \ , \\
& \\
\exp\left (- {2\mu}\,y^{1/\mu} \right )
& \hbox{(Nav04)} \ ,
\end{array}
\right.
\label{rhotildedark}
\\
g(y) &\!\!\!\!=\!\!\!\!& 
\left \{
\!\!\!\!
\begin{array}{ll}
\ln(y+1)-y/(y+1) & \!\!\!\!\hbox{(NFW)} \ ,
\\
& \\
2\,\left [
\sinh^{-1} \sqrt{y}
-\sqrt{y/(y\!+\!1)} \right ]&
\!\!\!\!\hbox{(JS--1.5)} \ , \\
& \\
{1\over2}\,\left (2\,\mu \right)^{1-3\,\mu}\,
\gamma \left [{3\,\mu},{2\,\mu}\,y^{1/\mu} \right ]
& \!\!\!\!\hbox{(Nav04)} \ ,
\end{array}
\right.
\label{gofc}
\end{eqnarray}
where $-\alpha$ is the inner slope,
$c$ is the concentration parameter (eq.~[\ref{cdef}]),
$a_h$ is the radius where the logarithmic slope is equal to $-2$ (NFW,
Nav04) or $-9/4$ (JS--1.5, for which $a_h/2$ is the radius where the slope is
$-2$), $\mu \simeq 6$ (appendix~\ref{appNav04}), and $\sinh^{-1} x =
\ln(x+\sqrt{x^2+1})$ for $x>0$.
In equation~(\ref{rhodark2}),
$M_v$ is the mass within the virial radius, defined
such that the mean total density within it
is $\Delta=200$ times the critical density of the Universe, $\rho_{\rm crit} =
3\,H_0^2/(8\,\pi G)$, yielding a virial radius (see \citealp{NFW97})
\begin{eqnarray}
r_v &=& \left ({2\,G\,M_v\over \Delta\,H_0^2}\right )^{1/3} \ ,\nonumber \\
&=& 163 \, h^{-1} \, {\rm kpc}
\,\left ({h\,M_v \over  10^{12}\,{\rm M_\odot}} \right )^{1/3} \ , \nonumber \\
&=& 206\,h_{70}^{-1} \, {\rm kpc}
\,\left ({h_{70} M_v \over  10^{12}\,{\rm M_\odot}} \right )^{1/3} \ ,
\label{rvir}
\end{eqnarray}
for $\Delta = 200$.
Equation~(\ref{gofc}) is derived by integrating equation~(\ref{rhodark2}) using
equation~(\ref{rhotildedark}), thus using equation~(\ref{MNav04}) for the
Nav04 model.
\cite{JS00} measured 
the
concentration parameter $c$ (eq.~[\ref{cdef}]) from their
$\Lambda$CDM (with cosmological density parameter 
$\Omega_m=0.3$ and dimensionless cosmological constant
$\Omega_\Lambda=0.7$) simulations, which can be fit by the relations
%
\begin{equation}
c \simeq \left\{
\begin{array}{ll}
\displaystyle
10.2 \,M_{12}^{-0.08}
& (\hbox{NFW}) \ , \\
& \\
\displaystyle
4.9 \,M_{12}^{-0.13}
& (\hbox{JS--1.5}) \ ,
\label{cJS}
\end{array}
\right.
\end{equation}
where
$M_{12} = {h M_v / 10^{12} {\rm M_\odot}}$.\footnote{The factor two
difference in NFW and JS--1.5 concentration parameters arises from the
different definition of scale radius $a_h$ is these two models.} 
In Appendix~\ref{appNav04}, we derive (eq.~[\ref{cNav04fit}]) the concentration
parameter for the Nav04 model:
\[
c = 8.1\,M_{12}^{-0.11-0.015\,\log M_{12}} \,
\]
which is
similar to the concentration parameters in equations~(\ref{cJS}).

X-ray data analyses \citep{Sato+00,WX00,LP02} based upon
equation~(\ref{MofrX}) conclude that the 
total mass of ellipticals, groups and clusters are indeed well fit by an NFW
model, but their concentration 
parameters increase much faster for decreasing masses. \citeauthor{Sato+00} 
give
\begin{equation}
c = 73 \,M_{12}^{-0.44}
\ ,
\label{cSato}
\end{equation}
while for the other two studies, the concentration parameter extrapolated to 
$M_v = 10^{12}\,h^{-1} {\rm M_\odot}$ ($M_{12}=1$) is slightly over 100 and the
slope is $-0.51$ in both cases.
Note that \citeauthor{JS00}, \citeauthor{Sato+00}, and 
Lloyd-Davies \& Ponman all define
their virial quantities at the radius where the mean density is 200 times the
critical value, while Wu \& Xue define their virial radius where the mean
density is 100 times the critical value, in accordance with the spherical top
hat infall model for $\Omega_m =0.3, \Omega_\Lambda=0.7$.
Interestingly, in their cosmological $N$-body simulations,
\cite{Bullock+01} confirm high concentration parameters for the mass
distribution within \emph{subhaloes}
of haloes (which may correspond to elliptical galaxies within clusters), with
a steeper dependence of concentration on mass:
$d\ln c/d\ln M \approx -0.3$.

The cumulative mass of the dark models used here can all be written
\begin{eqnarray}
M(r) &=& M_v\,\widetilde M(r/a_h) \label{Mdarkofr} \ ,\\
\widetilde M(y) &=& {g(y) \over g(c) } \ ,
\label{mtildedark}
\end{eqnarray}
where $g$ is given in equation~(\ref{gofc}).\footnote{Note that the
definition of $g$ is the inverse of the definition 
of $g$ given by \cite{LM01} for the NFW model.}

\section{Results}
\label{results}
\subsection{Local mass-to-light ratios}
\label{totalmass}

The simplest check that cuspy $\Lambda$CDM models represent the \emph{total}
mass 
distribution of elliptical galaxies, i.e. the gravitational potential, is
by checking that the mass at all radii is greater than the known contribution
from stars.

Given equations~(\ref{cdef}), (\ref{nuofr}), (\ref{nuofx}), (\ref{nu1}),
(\ref{rhodark2}), 
(\ref{rhotildedark}), and
(\ref{gofc}),
the \emph{local} mass-to-light ratio is
\begin{eqnarray}
\Upsilon(r) &=& {\rho(r) \over \ell(r)}
= {\rho_1\over \ell_1}\,{\widetilde \rho(r/a_h) \over \widetilde \ell(r/a_S)} \
,\nonumber \\
&=& {m\,\Gamma[(3-p)m]\over g(c)}\,{M_v\over
L}\,\eta^{3-\alpha} \nonumber \\
&\mbox{}& \times \,x^{p-\alpha} \,(1+\eta x)^{\alpha-3} \exp\left
(x^{1/m}\right ) \ ,
\label{moverl}
\end{eqnarray}
where the second equality is restricted to the NFW and JS--1.5 models, with
$x=r/a_S$, $\eta = a_S/a_h$, and $\alpha=1$ (NFW) or 3/2 (JS--1.5).

\begin{figure}
\includegraphics[width=\hsize]{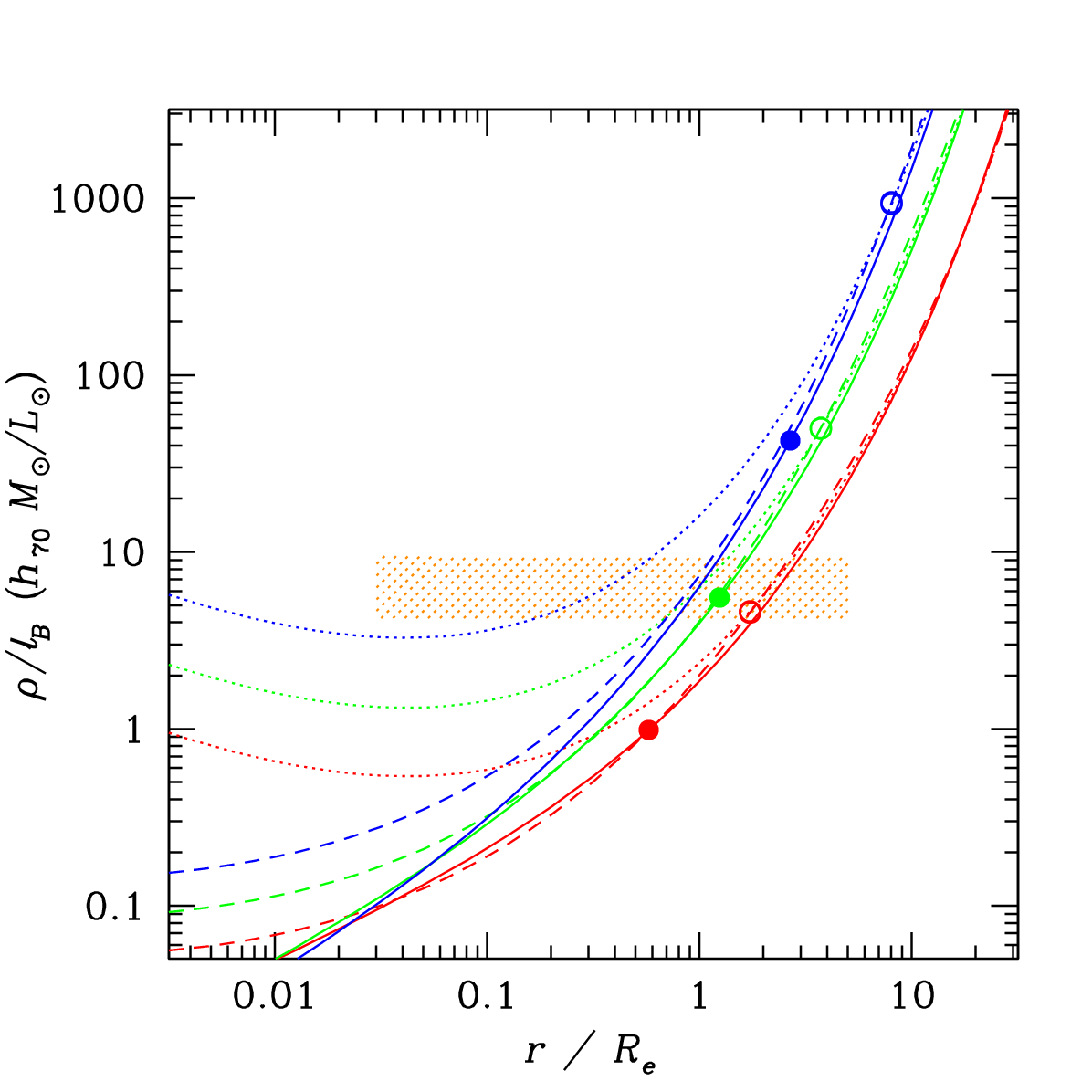}
\caption{Local mass-to-light ratio profiles for the S\'ersic distribution
of luminosity $L_B = L_B^* = 1.88 \times
10^{10}\,h_{70}^{-2}\,{\rm L_\odot}$, with
$m=3.1$ (eq.~[\ref{mfit}]) and $R_e = 3.2 \, h_{70}^{-1} \, \rm kpc$
(eq.~[\ref{Refit}]),
and NFW (\emph{dashed curves}), JS--1.5 (\emph{dotted
curves}), and Nav04 (\emph{solid curves}) 
total mass distributions with $\log h_{70}^{-1}\,M_v = 11.87$, 12.87, and
13.87 (mass-to-light ratios $\Upsilon_B = 39, 390, 3900$, i.e.
0.1, 1, and 10 times the universal value of eq.~[\ref{mluniv}]),  
increasing upwards (\emph{red}, \emph{green}, and \emph{blue}
in the electronic version of the Journal).
The \emph{open circles} represent $0.03\,r_v$, the minimum radius where the
NFW and JS--1.5 models are accurate in representing the density profiles in
structures found in cosmological $N$-body
simulations, while the \emph{filled circles} represent $0.01\,r_v$, which is
the analogous radius for the Nav04 model.
The curves are obtained with equation~(\ref{moverl}) using
equations~(\ref{cdef}),
(\ref{pofm}), (\ref{Re}), (\ref{bofm}), (\ref{Refit}), 
(\ref{gofc}), (\ref{rvir}), and (\ref{cJS}).
The \emph{shaded region} (\emph{orange} in the electronic version of the
journal) shows the mass-to-light
ratio of the stellar populations of elliptical galaxies (in the range of radii
where the S\'ersic fits to the surface brightness profile are believed to be
good, see Sec.~\ref{scalings}).
}
\label{mlofrups0}
\end{figure}

Fig.~\ref{mlofrups0} shows the local $M/L$ profile for 
the S\'ersic model with luminosity $L_B = L_* = 1.88 \times 
10^{10}\,h_{70}^{-2}\,{\rm L_\odot}$ and 
our three adopted mass models with
masses within the virial radius of 
$\log h_{70}^{-1}\,M_v = 11.87$, 
12.87, and 13.87, i.e. for mass-to-light ratios 0.1, 1, and 10 times
 the universal value (eq.~[\ref{mluniv}]).
The shaded region in Fig.~2 indicates the mass-to-light ratios inferred
from stellar population synthesis (see Sec.~\ref{scalings}), in the range
where the surface brightness profile is well known (and where the
S\'ersic deprojection is considered accurate, see Sec.~\ref{dislight}).
Fig.~\ref{mlofrups0} indicates that,  at the lower limit of its spatial
resolution ($0.01\,r_v \simeq 0.6\,R_e$), 
the low mass ($\Upsilon_B = 39$) Nav04 model produces a lower mass-to-light
ratio than stellar, reaching values as low as unity, inconsistent with all
estimates of the stellar mass-to-light ratios of old (and red, as observed)
stellar populations.

For the models with higher virial cumulative mass-to-light ratio, the local
mass-to-light ratios are consistent with the expected stellar
mass-to-light ratios in the range where these dark matter models are
measured.
However, simple inwards extrapolations of these mass models rapidly lead to
sub-stellar mass-to-light ratios.
For example, for the universal virial mass-to-light ratio ($\Upsilon_B =
390$), the Nav04 model is consistent with the stellar mass-to-light ratio 
at the limit of its resolution, $r = 0.01\,r_v \simeq 1.4\,R_e$.
But if one wishes to ensure that 
the local mass-to-light ratio at lower radii remains higher than the
stellar value, one then requires the
mass model to have a density slope equal to that of a S\'ersic model.
{}From equations~(\ref{nuofr}), (\ref{nuofx}) and (\ref{pofm}),
the slope of the deprojected S\'ersic profile is
$-p-b/m (r/R_e)^{1/m}$.
For $m=3.1$, one gets a slope of $-2.9$ at $1.4\,R_e$, and of $-1.4$ at
$0.03\,R_e$. In comparison, for elliptical galaxies with
$h M=10^{12} M_\odot$, for which the Nav04 resolution limit is 
$r = 0.01\,r_v = 1.4\,R_e = 0.08\,r_{-2}$
(eq.~[\ref{cNav04fit}]), then
according to equation~(\ref{Nav04slope}),
the Nav04 model has a slope of $-1.3$,
as also inferred from fig. 3 of \citeauthor{Navarro+04}. While the density
profile of 
the Nav04 model becomes shallower with decreasing radius (as do the density
profiles of the NFW
and JS--1.5 models), one must, on the contrary, assume that the profile
suddenly steepens at $r < R_e$. The same argument applies for the NFW model.
Hence, the NFW and Nav04 models cannot represent the total mass.

The cuspier nature of the JS--1.5 model makes it less sensitive to this
criterion. Nevertheless, Fig.~\ref{mlofrups0} indicates that a slope
steeper than $-1.5$ is required to keep the mass-to-light ratio above the
stellar value.
Therefore, \emph{unless 
there is a sharp break in the density profiles inside the
resolution radius to a slope that is considerably steeper than $-3/2$, any
reasonable extrapolation of the 
mass-to-light ratio curves will lead to values lower than stellar}.

Do the high-concentration NFW models found by \cite{Sato+00} for the
total matter distribution also produce abnormally low mass-to-light ratios?
For the three elliptical galaxies studied by
\citeauthor{Sato+00}: 
{NGC~1399},
{NGC~3923}, and
{NGC~4636}, we find from the {\it LEDA} database
extinction-corrected total blue
magnitudes of 10.25, 10.38 and 10.22, respectively, and distance moduli from
the surface brightness fluctuation study of \cite{Tonry+01}, corrected by
$-0.16$ magnitude  (see
\citealp{Jensen+03}) for the newer Cepheid distance normalization of
\cite{Freedman+01}, of 31.34, 31.64
and 30.67, respectively, hence $L_B = 4.1$, 4.8 and $2.3 \times 10^{10}
{\rm L_\odot}$, respectively.

\begin{figure}
\includegraphics[width=\hsize]{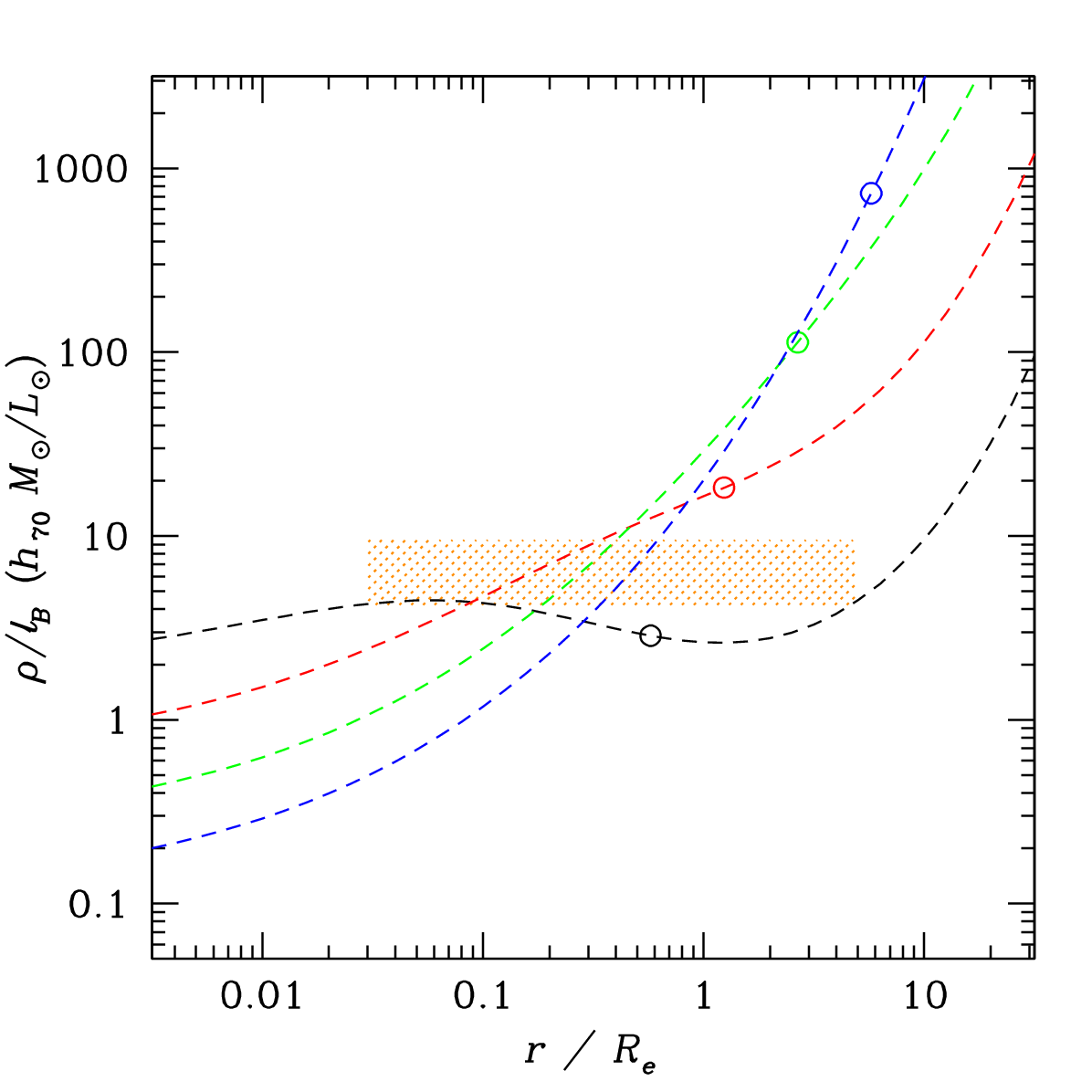}
\caption{Same as Fig.~\ref{mlofrups0} for high concentration parameters
obtained from X-ray analyses by Sato et al. (2000) for the NFW model and for
$L_B = 4.1\times10^{10} {\rm L_\odot}$ (the median luminosity of the 3
elliptical 
galaxies in their sample), yielding $m=3.7$ from equation~(\ref{mfit}) and
$R_e = 5.8 \, h_{70}^{-1} \, \rm kpc$ from equation~(\ref{Refit}).
The log masses are now 11.20, 12.20, 13.20 and 14.20
(respectively 
corresponding to $\Upsilon = 3.9, 39, 390$, and 3900, i.e. 
0.01, 0.1, 1, and 10 times the universal mass-to-light ratio of
equation~(\ref{mluniv}), and respectively 
\emph{black}, 
\emph{red}, \emph{green}, and \emph{blue} 
in the electronic version of the Journal)
going downwards at small radii.
}
\label{mlbigc}
\end{figure}
We plot in Fig.~\ref{mlbigc} the local mass-to-luminosity ratios for
NFW models with 
the high concentration parameters of Sato
et al. (eq.~[\ref{cSato}]) for the median blue luminosity
$L_B=4.1\times10^{10}\,{\rm L_\odot}$.
%
%

%
%
The local 
mass-to-light ratio profiles are increased in the inner regions,
relative to the analogous profiles for the NFW model with the low
concentration parameters as found in the cosmological $N$-body simulations.
This is
especially the case for low
masses, which have the most centrally concentrated mass-density profiles.
Nevertheless, 
the local mass-to-light ratio found with the high concentration parameter of
\citeauthor{Sato+00} is below the stellar value at all radii below $0.08\,R_e$,
for the very large range of masses considered.
But here,
one could imagine dark matter profiles with somewhat steeper inner slopes
well within 3\% of $r_v$
that would produce local mass-to-light ratios in excess of the stellar value
at all radii where the S\'ersic model is a good fit to the surface brightness
profile of ellipticals.
The same conclusions are reached for the even higher concentration parameters
extrapolated to the masses of ellipticals from the relations of 
\cite{WX00} and Lloyd-Davies \& Ponman.

To summarise, \emph{the low-concentration matter distributions
found in cosmological $N$-body simulations produce local
mass-to-light ratios lower than expected for stellar populations, unless
there is a sharp steepening 
in the slopes of the mass density profiles just within an
effective radius,
while high-concentration NFW profiles would produce large enough local
mass-to-light ratios if their inner slopes were slightly steeper.}

These conclusions are unchanged if we vary the effective radius or the
S\'ersic shape ($m$) by a factor of 2.

\subsection{Velocity dispersions}

Another
way to check the compatibility of NFW, JS--1.5 and Nav04
potentials with the 
S\'ersic luminosity profiles of ellipticals is to compute the central
stellar line-of-sight
velocity dispersion, averaged within an aperture or a slit.
In appendix~\ref{appJeans}, we re-derive the stellar
radial and line-of-sight velocity 
dispersion profiles, assuming isotropy, 
and derive for the first time the velocity dispersion
profiles averaged in circular apertures and thin slits
in terms of single quadratures of the tracer
density and total mass profile.\footnote{One can find in the literature the
use of triple integrals to compute aperture velocity dispersions (e.g.,
\citealp{BSD03}) whereas the single integrals given in
appendix~\ref{appJeans} are much simpler to evaluate.}

\begin{figure*}
\includegraphics[width=0.7\hsize]{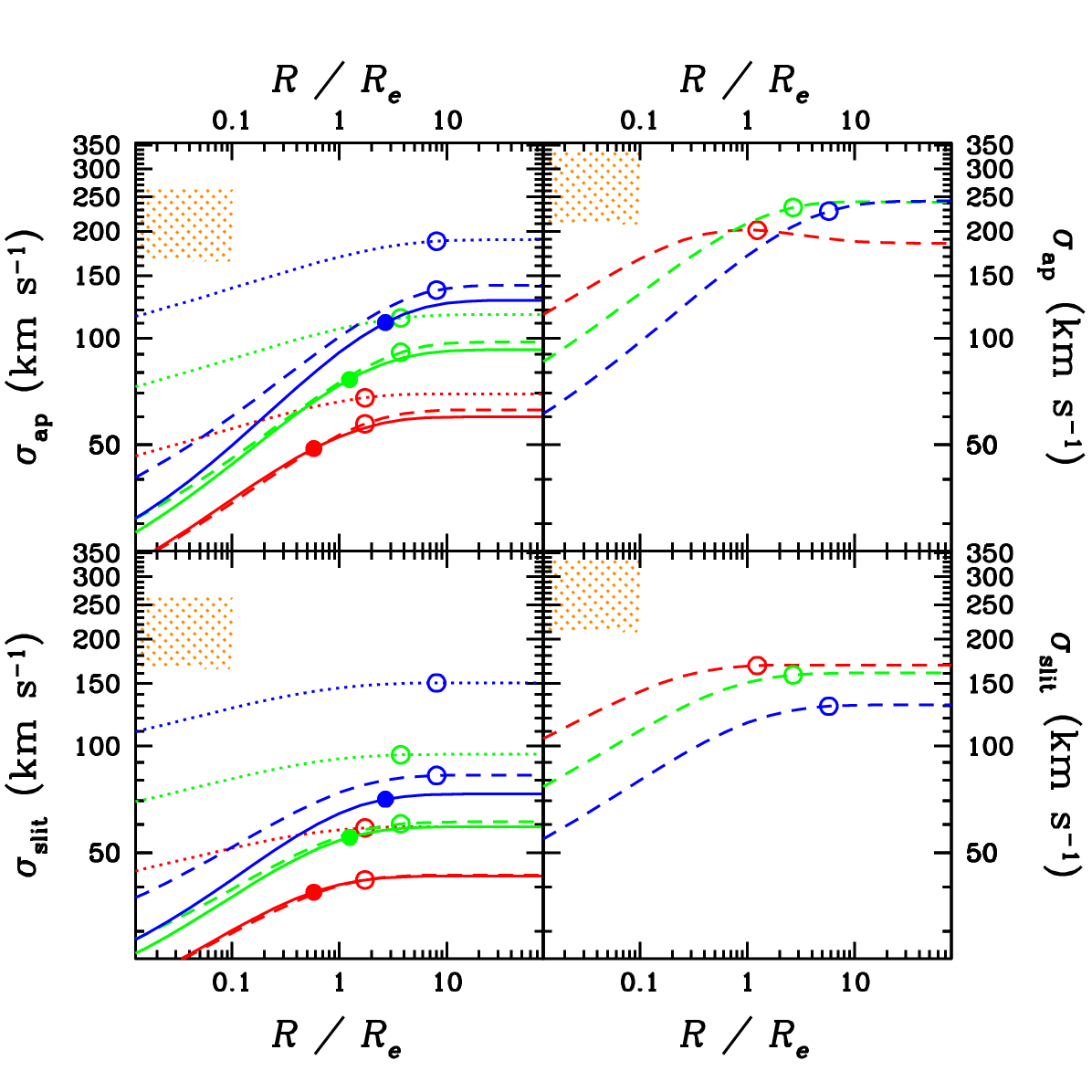}
\caption{Stellar isotropic velocity dispersions, averaged over circular
apertures of projected radius $R$ 
(\emph{top}, eq.~[\ref{sigapiso}]) or thin slits of half-projected size 
$R$ (\emph{bottom}, eq.~[\ref{sigslitiso}]) for S\'ersic luminosity profiles
and CDM dark matter profiles: NFW (\emph{dashed curves}), JS--1.5
(\emph{dotted curves}) and Nav04 (\emph{solid curves}).
\emph{Left}: low concentration parameters from cosmological simulations 
with total luminosity
$L_B = L_B^* = 1.88 \times 10^{10}\,h_{70}^{-2} {\rm L_\odot}$ 
(corresponding to S\'ersic parameters
$m=3.1$ and $R_e = 3.2 \, h_{70}^{-1} \, \rm
kpc$).
\emph{Right}: high concentration parameters from X-ray observations
(eq.~[\ref{cSato}], NFW model only), with total luminosity 
$4.1\times 10^{10}\,h_{70}^{-2} {\rm L_\odot}$ 
($m=3.7$ and $R_e = 5.8 \, h_{70}^{-1} \, \rm
kpc$).
Equations~(\ref{pofm}), (\ref{L3tilde}), (\ref{rvir}), 
(\ref{gofc}), (\ref{Mdarkofr}) and (\ref{mtildedark})
are used.
%
The cumulative mass-to-light ratios within the virial radius $r_{100}$ are
39, 390 and 3900
increasing upwards in the \emph{left plots} and downwards in the \emph{right
plots}
(\emph{red}, \emph{green}, and
\emph{blue} in the electronic version of the Journal).
The \emph{shaded regions} are the central observed velocity dispersions
expected from
the Faber-Jackson (\citeyear{FJ76}) scaling relation, recalibrated by 
\citeauthor{BZB96} (\citeyear{BZB96}, eq.~[\ref{fjbender}]).
The \emph{open circles} represent $0.03\,r_{200}$, the minimum radius where the
NFW and JS--1.5 models are accurate in representing the density profiles in
structures found in cosmological $N$-body
simulations, while the \emph{filled circles} represent $0.01\,r_{200}$, which
is the analogous radius for the Nav04 model.
}
\label{sigapslitUps0}
\end{figure*}

Fig.~\ref{sigapslitUps0} shows the resulting aperture and slit-averaged
velocity dispersion profiles for S\'ersic tracers and NFW, JS--1.5 and Nav04 
mass
profiles. 
%
%
The figure clearly shows that, if the \emph{total} matter is represented by
an NFW, 
JS--1.5, or Nav04 density models, with low concentration parameters as found in
cosmological simulations,
then the central aperture and slit velocity dispersions would be much
smaller than observed.
Indeed, the aperture and slit-averaged velocity dispersions both
obey $\sigma_v < 113 \, \rm km
\, s^{-1}$ for $R < 0.1\,R_e$ (where the equality is reached for the more
favourable JS--1.5 model), whereas one expects $\sigma_v = 208 \, \rm km
\, s^{-1}$, from 
the Faber-Jackson scaling relation, recalibrated by Bender et
al. (eq.~[\ref{fjbender}]).
For the most realistic Nav04 model, the velocity dispersions are
3.5 times too low for $R < 0.1\,R_e$.

One might worry that this conclusion is reached through heavy extrapolation
of the dark matter models at small radii. However, it is again difficult to
imagine a dark matter model that would produce aperture or slit velocity
dispersions that rise sharply (factor of 3 for the Nav04 model from 2 to
$0.05\,R_e$) at increasingly small radii. Given the difference between the
mass models of inner slope $-1$ (NFW) and $-3/2$ (JS--1.5),
Fig.~\ref{sigapslitUps0} suggests that an inner slope $< -2$ is required
for the mass distribution to recover the high observed\
central velocity dispersions.

Note that the slit-averaged velocity dispersions are slightly smaller than the
aperture-averaged ones, as is indeed expected given that, for the luminosity
and mass models
considered here,
the line-of-sight velocity dispersions increase
with radius for $R < R_e$, and the slit velocity dispersions are less weighted
to outer radii than are the aperture velocity dispersions.
Hence, the slit velocity dispersions are in principle
more constraining against a global
NFW, JS--1.5 or Nav04 potential.
However, our quasi-analytical expressions for the slit velocity dispersion
neglect the effects of seeing and are thus of little use at small radii.
Moreover, at large radii observers subdivide their slit into rectangular
bins, whose modelling is beyond the scope of this paper.
We therefore, will focus on aperture velocity dispersions.

\begin{figure*}
\includegraphics[width=0.7\hsize]{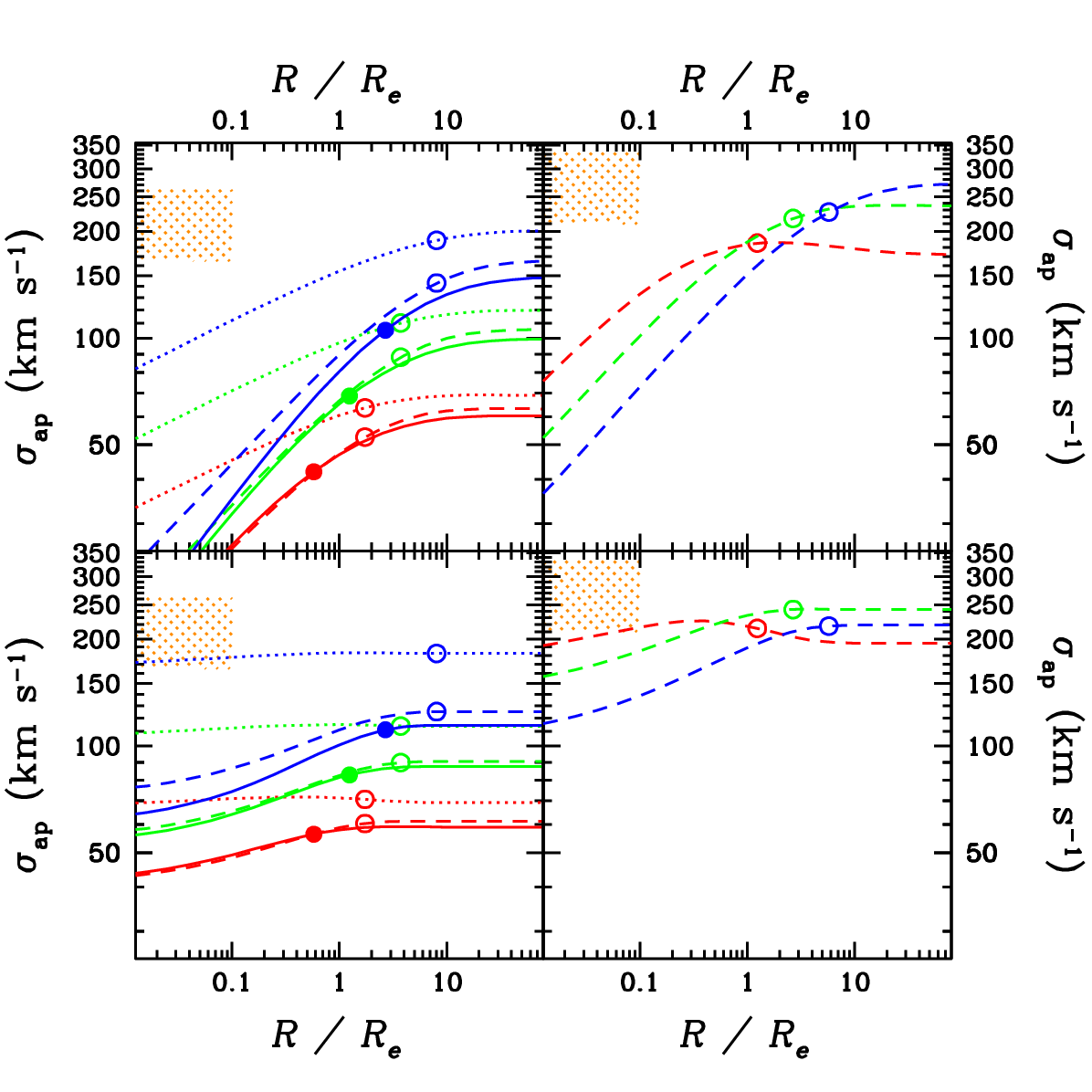}
\caption{Same as the top plots of Fig.~\ref{sigapslitUps0}, 
for S\'ersic shape parameter
equal to twice ($m=6.2$ and 7.4, \emph{upper left and right plots},
respectively) or half ($m=1.55$ and 1.85,
\emph{lower left and right plots}, respectively) of the values
derived from eq.~(\ref{mfit}).
}
\label{sigapslitmfac}
\end{figure*}
The high concentration parameters derived from X-ray observations produce
higher central aperture and slit velocity dispersions, but these velocity
dispersions are still well
below the prediction of the Faber-Jackson relation: the
NFW model produces slit velocity dispersions that are 1.5 times too low for
$R < 0.1\,R_e$ for all mass-to-light ratios below 3900 (10 times the
universal value of eq.~[\ref{mluniv}]).

As mentioned in Sec.~\ref{dislight}, the S\'ersic shape-luminosity relation
is probably uncertain 
by a factor of perhaps 2 (Fig.~\ref{figcor}).
Fig.~\ref{sigapslitmfac} shows the dispersion profiles when the S\'ersic
shape $m$ is taken to be twice (\emph{top plots}) or half (\emph{bottom
plots}) the value given in equation~(\ref{mfit}).
%
The higher S\'ersic parameters favoured
by \cite{BJ98} and \cite{PS97} produce even lower
velocity dispersions, and are thus even more constraining against the idea
that the global potential of elliptical galaxies can be that of the
$\Lambda$CDM simulations.
And if one adopts a S\'ersic shape parameter as low as
$m=1.35$, as derived from the \citeauthor{Marquez+00} data 
for the galaxies in Abell
85,
one derives central aperture velocity dispersions that are almost large
enough, but this requires the enormous mass-to-light ratio $\Upsilon_B =
3900$, i.e. a total mass of over $3\times 10^{13}\,h_{70}^{-1} {\rm M_\odot}$,
which is larger than that of a rich group.
For the case of the high concentration parameter found in the X-rays, the
aperture velocity dispersions are still too small, for all reasonable S\'ersic
parameters.
Moreover, 
given that the de Vaucouleurs surface brightness profile ($m=4$) was derived
for ellipticals of typically 0.5 to $1\,L_*$, we are very doubtful that
a S\'ersic parameter as low as $m=1.27$ is 
representative of fairly luminous elliptical
galaxies.

Therefore,
\emph{the internal kinematics of elliptical galaxies
appear to be 
inconsistent with the total mass distribution as found in  cosmological
$N$-body simulations, even with the steep inner
slope of $-3/2$ or the high concentration parameters found by X-ray
observers}. 
%


\subsection{Large concentration mass models as seen in X-rays}

If the NFW model cannot represent the total density profile of elliptical
galaxies, how can one explain that several X-ray studies converged on a high
concentration parameter NFW total density profile?

\begin{figure}
\includegraphics[width=\hsize]{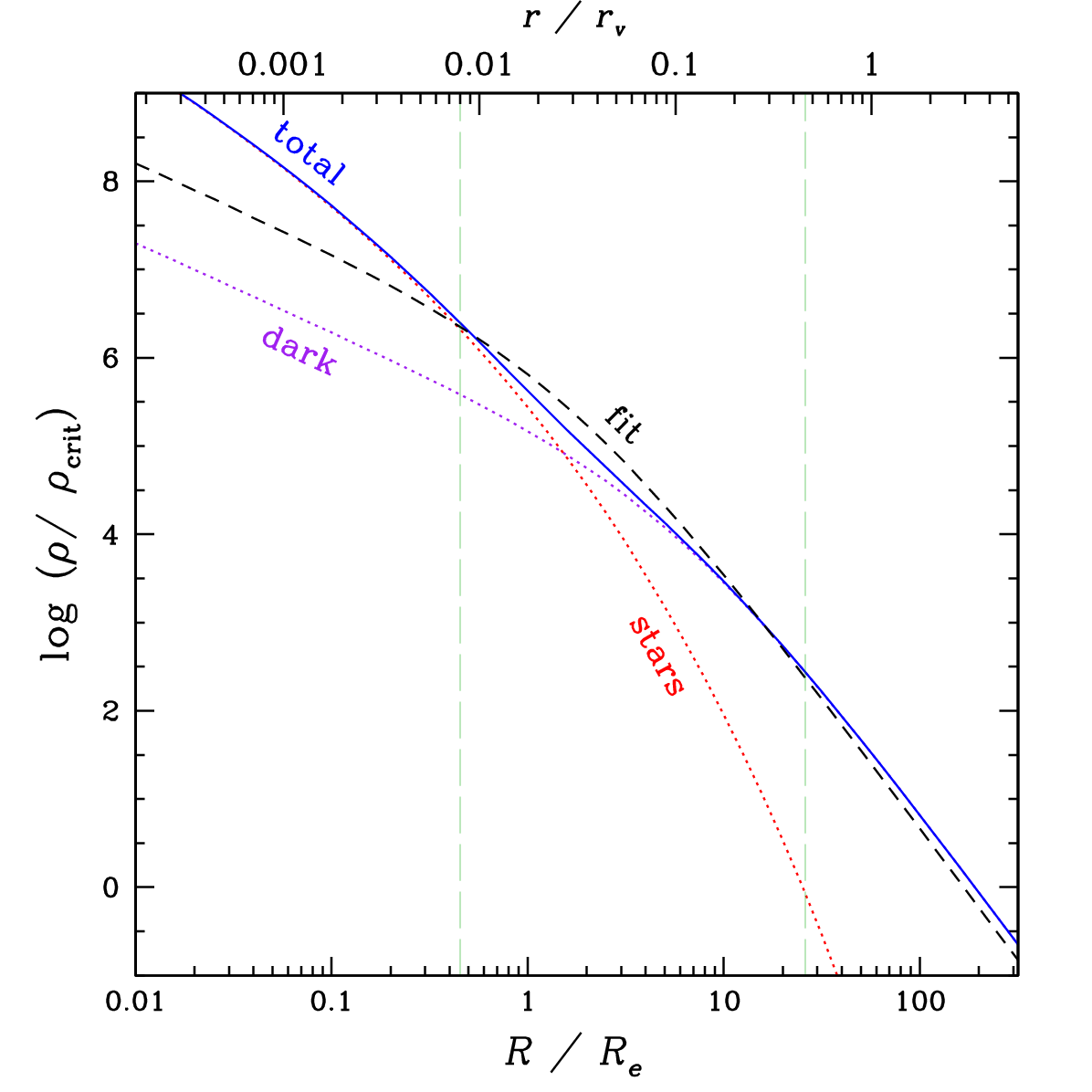}
\caption{Density profiles: stars (\emph{dotted curve}, 
$L_B = 4.1\times10^{10}\,h_{70}^{-2} {\rm L_\odot}$,
$m=3.7$, $R_e=5.8 \, h_{70}^{-1} \, \rm
kpc$, $\Upsilon_{*,B}=8$), NFW-dark matter (\emph{dotted curve}, $M_v =
3.8\times 
10^{12}\,h_{70}^{-1} {\rm M_\odot}$, 
$c=9.5$) and
total (\emph{solid curve}) and the NFW fit to the total density
(\emph{dashed curve}), with $c=35$.
The vertical lines delimit the region of the fit (see text).
The top abscissa are relative to $r_{200}$.
}
\label{fitrhototnfw}
\end{figure}

The answer may be seen in
Fig.~\ref{fitrhototnfw}, which shows the stellar (eqs.~[\ref{nuofr}],
[\ref{nuofx}], and [\ref{Ltot}])
and dark matter (eqs.~[\ref{rhodark2}], [\ref{rhotildedark}], [\ref{gofc}],
[\ref{rvir}], and [\ref{cJS}])
density profiles for the median luminosity ($L_B = 4.1 \,h_{70}^{-2} {\rm
L_\odot}$) of the 3 elliptical galaxies in
the sample of \cite{Sato+00}, with effective radius and S\'ersic shape
parameter taken from equations~(\ref{Refit}) and (\ref{mfit}), respectively,  
and for the mass that \citeauthor{Sato+00} derived within the virial
radius ($M_v=3.6 \,h_{70}^{-1} \times 10^{12}\,{\rm M_\odot}$, 
leading to $\Upsilon_B = 75$).
Here, we
fix the virial radius ($r_v=316\, h_{70}^{-1} \,\rm kpc$ from
eq.~[\ref{rvir}]), and we fit the best 
NFW model to the density profile, within the limits shown as vertical lines
in Fig.~\ref{fitrhototnfw} 
(26 arcsec, set by the PSF of the {\it ASCA} telescope, and 25
arcmin, which is half the field of view of the {\it GIS} instrument on {\it
ASCA}). With the median distance modulus of 31.34 (see Sect~\ref{totalmass}),
these spatial limits respectively 
correspond to 2.3 and $129 \, h_{70}^{-1} \, \rm kpc$,
i.e., 0.0074 and 0.41 virial radii.
Since \citeauthor{Sato+00} fit their \emph{mass} profiles with an NFW mass
profile, we minimize the square differences in log mass 
over
equally-spaced log radii.

We find 
a high concentration parameter: $c=35$, although not as high as $c=49$,
inferred from
the general concentration vs. mass relation of \citeauthor{Sato+00}
(eq.~[\ref{cSato}]), and only half the value actually found by
\citeauthor{Sato+00} for these 3 galaxies (which 
were positive outliers on their
global distribution of concentration as a function of mass).
Moreover, using $\Upsilon_{*,B} = 5$ instead of 8, the best-fitting 
concentration
parameter (with the same quality fit as in
Fig.~\ref{fitrhototnfw}) is reduced to $c=26$, still much higher than $c=9.5$
inferred from the cosmological simulations of \citeauthor{JS00}
(eq.~[\ref{cJS}]) for NFW models with the mass inferred by
\citeauthor{Sato+00}. 
Note that the best fit, although adequate, is not superb, because the sum of
the stellar and dark matter components presents an inflexion point at the
radius where the two components contribute equally ($R \simeq 1.1$ and
$1.4\,R_e$ for $\Upsilon_{*,B} = 5$ and 8, respectively), in 
contrast with the NFW model, whose slope steepens continuously with
increasing radius.

Hence, Fig.~\ref{fitrhototnfw} tells us that
\emph{fitting an NFW model to the total density represented by the
sum of the NFW dark matter component and the less extended 
S\'ersic stellar component can produce very high concentration parameters,
but the fits are not excellent}.


\section{Summary and Discussion}

Given what we know about the surface brightness profiles of elliptical
galaxies, one can simply conclude that the observations of
elliptical galaxies cannot allow NFW total density profiles with the fairly
low concentration parameters as found in cosmological $N$-body simulations: 
otherwise one
would end up with mass-to-light ratios far below stellar within an effective
radius, and aperture and
slit velocity dispersions would be far 
below the predictions from the isotropic Jeans
equations.
Trying a density profile with a steeper inner slope as advocated by
\cite{FM97} and \cite{Moore+99}, using the model of \cite{JS00}, does not
help, nor does the recent model by \cite{Navarro+04}, nor probably the very
similar mass model proposed by Stoehr and colleagues \citep{SWTS02,Stoehr05}.

This conclusion of non-conformity of $\Lambda$CDM mass models with the
S\'ersic luminosity profile of elliptical galaxies is based upon the
extrapolation of these mass models to radii smaller than the spatial
resolution of the simulations.
Nevertheless, the only way to bring the simulated mass models in conformity
with the observations is for the slopes of the mass density profiles to
sharply break around the effective radius from values $\simeq -1.3$ just
outside $R_e$ to steeper values $<-2$ inside $R_e$, and it is hard to imagine
that this sharp break in slope happens at the precise spatial resolution
limit of present-day cosmological $N$-body simulations.
This conclusion is independent of any assumed value for the stellar
mass-to-light ratio (as are the curves in Figs.~\ref{mlofrups0} and
\ref{mlbigc}, as well as all of Figs.~\ref{sigapslitUps0} and 
\ref{sigapslitmfac}). 

However, allowing for ten times larger concentration parameters, as motivated
by several X-ray studies, makes the mass-to-light ratios and aperture/slit
velocity dispersions almost in line with the observations, but not quite.

Assuming no sharp break in the slope of the total mass density profile to
steeper than $-2$ inside $R_e$,
we are left with an inconsistency of too 
low mass-to-light ratios and velocity dispersions of the inner regions.
This inconsistency 
can
disappear if we add more mass in these inner regions.

The simplest way to add more mass inside an NFW-like mass distribution is to
postulate that this distribution only applies to the dark matter component
and that the baryons, essentially stars, dominate the mass profile in the
inner regions, as their S\'ersic distribution has a steeper inner slope than
the NFW-like models. In the following paper \citep{ML05b}, we study in more
detail  
multi-component models of elliptical galaxies.

Moreover, 
during the formation process of ellipticals, the NFW-like dark matter
should dynamically
respond to the dominant inner baryons. This process is often studied through the
approximation of adiabatic contraction \citep{BFFP86}.
However, the dark matter response to the dominant presence of the stellar
component will diminish as its local density approaches that of the stars.
Therefore, the dark matter cannot dominate the stars inside, although it
could conceivably reach non-negligible densities relative to 
the stellar component.
%
%
In any event,
\emph{as long as the inner slope of the dark matter density profile in
elliptical galaxies is
shallower than
$-3/2$, it should be difficult to constrain its precise value 
in elliptical galaxies}.

The very different mass distribution for the baryonic (mainly stellar)
component has important ramifications.
It suggests that present-day
elliptical galaxies
have not recently formed through rapid 
collapse, during which violent relaxation should lead to an
efficient mixing of the baryon and dark matter components, especially if
their initial distributions were similar (in the linear regime of small
density fluctuations).

One alternative is that ellipticals form early by collapse, and the
dissipative nature of the baryons slowly leads to a segregation with the dark
matter, which in turn slowly responds to the contracting baryons by
contracting itself.
\cite{GKKN04} and \cite{Dekel+05} find dark matter density profiles scaling
as $r^{-2}$ over a very large range of radii in their haloes and
elliptical-galaxy-like merger
remnants of their cosmological and two-galaxy $N$-body simulations,
respectively.
Although these findings may be inconsistent with several observational
constraints of elliptical and spiral galaxies (perhaps because of
insufficient feedback from star formation and AGN that pumps energy into
the inner baryons, which should in turn puff up the inner dark matter
distribution), they confirm our finding that the total mass density profile
cannot be as shallow as predicted by the dissipationless 
$\Lambda$CDM mass models.

The other important alternative is the formation of elliptical galaxies
(i.e. their morphologies,
not their stars, which are thought to have formed earlier) 
through major mergers of
spiral galaxies. Since spiral galaxies are initially gas-rich and form by
dissipational collapse, a baryon / dark matter segregation  sets in as the
baryons settle in a disk, while the dark matter extends further out.
When spiral galaxies of comparable mass merge into ellipticals, the 
baryons, which are more tightly bound than the dark matter particles, will
end up in the inner regions of the elliptical merger remnant
(e.g., \citealp*{DDH03}; \citealp{Dekel+05}).

The second conclusion of this work is that considering a two component model
for elliptical galaxies, summing up the mass density
profiles of the S\'ersic stellar component
(with a reasonable mass-to-light ratio) and a dark matter
component as seen in the $\Lambda$CDM dissipationless cosmological $N$-body
simulations, one finds a total mass density profile that resembles an NFW
model, with a concentration almost as high as deduced from X-ray
observations, much higher than in $\Lambda$CDM haloes.
Admittedly, the fit is not excellent.

This high concentration NFW total mass distribution is caused by the more
extended nature of $\Lambda$CDM haloes in comparison with the stellar
distribution of elliptical galaxies.
It would be good if the X-ray observers attempt to fit NFW or better
Nav04 models to the dark matter component of their elliptical galaxies or
groups and not to the total mass density
profile. 

Note that recent modelling by \cite{PA05} of their {\it XMM-Newton} 
X-ray observations of poor and rich clusters of
galaxies leads to the conclusion that the total mass density
profile is consistent with NFW models of normal concentration, with the same
shallow dependence on mass as seen in the cosmological $N$-body simulations.
This result is
in stark contrast with the strong dependence of concentration with mass found
by \cite{Sato+00} from ellipticals to poor clusters to rich clusters.

One may venture that elliptical galaxies have a more prominent baryonic
component in their inner 
regions than do  groups and
clusters of galaxies, and will thus, by superposition of S\'ersic and
$\Lambda$CDM components, appear more concentrated when fitted with single
NFW models, than expected from the general trend of groups or clusters.
However, \citeauthor{Sato+00} find a continuous power-law trend for the mass
dependence of their concentration parameters. In other words,
\citeauthor{Sato+00} find moderately high concentration parameters for 
groups of galaxies, as do \cite{WX00} and \cite{KJP04},
suggesting an important baryonic contribution in their inner regions, albeit
less dominant than in elliptical galaxies.

Interestingly, our best NFW model for the dark component of the 2-component
model we try for the 
galaxies observed by \citeauthor{Sato+00}, has a concentration,
$c\simeq 26$ to 35, that lies in between the very high
value  of \citeauthor{Sato+00}, $c \simeq 49$, and the low value expected from
cosmological $N$-body simulations for objects of that virial mass, $c\simeq
9.5$. 
Additional measurements of concentration from X-ray observations of
ellipticals should greatly clarify this issue.

In the following paper \citep{ML05b}, we investigate in more detail a
4-component model, including stars, dark matter, hot gas and a central black
hole, and ask whether the observations of ellipticals require cuspy cores,
and to which accuracy one can measure the total mass within the virial radius
through kinematic observations.

















\section*{Acknowledgments}

We are happy to thank 
Bernard Fort, 
Daniel Gerbal,
Gastao Lima Neto,
Jochen Liske,
Alain Mazure,
Thanu Padmanabhan,
Jose Maria Rozas,
and
Felix Stoehr
for useful
comments.
We also gratefully acknowledge
Fumie Akimoto for providing us with details on the X-ray observations 
of the elliptical galaxies
studied by \cite{Sato+00}, 
and Yi Peng Jing for providing us with the JS--1.5 concentration
parameter from his simulations in digital form.
We warmly thank the referee, Aaron Romanowsky, for his numerous insightful
comments.
This research made use of the {\it LEDA} galaxy 
catalogue of the {\it HyperLEDA}
galaxy database ({\tt http://www.leda.univ-lyon1.fr}).
EL\L \
acknowledges hospitality of Institut d'Astrophysique de Paris, where
part of this work was done, while GAM benefited in turn from the hospitality
of the Copernicus Center in Warsaw.
This research was partially supported by the
Polish Ministry of Scientific Research and Information Technology
under grant
1P03D02726
as well as the Jumelage program Astronomie France Pologne of
CNRS/PAN.

\appendix
\onecolumn

\section{Concentration, central density and outside mass 
versus virial mass for the Navarro et al. (2004) model}
\label{appNav04}
\cite{Navarro+04} have recently shown that the density profiles of haloes
in cosmological $N$-body simulations begun with a $\Lambda$CDM
spectrum can be fit to high precision to 
\begin{equation}
\rho(r) = \rho_{-2}\,\exp (2 \mu)\,\exp\left [-2 \mu \,\left ({r\over r_{-2}}
\right )^{1/\mu} \right ] \ ,
\label{rhonav04}
\end{equation}
where our $\mu$ is the inverse of their $\alpha$ and where $\rho{_2}$ is the
local mass density at the radius $r_{-2}$ 
where the logarithmic slope of the density
is $-2$.

The enclosed mass of the density profile of equation~(\ref{rhonav04}) is
\begin{equation}
M(r) = 4\pi\,\rho_{-2}\,
r_{-2}^3\,2^{-3\mu}\,\mu^{1-3\mu}\,\exp(2\,\mu)\,\gamma\left [3\mu,2\mu
\left ({r\over r_{-2}}\right )^{1/\mu} \right ] \ ,
\label{MNav04}
\end{equation}
where $\gamma(a,x) = \int_0^x t^{a-1}\,\exp(-t)\,dt$ 
is the incomplete gamma function.
\cite{Navarro+04} provide a table of values of $1/\mu$, which for giant 
galaxy mass
objects yield a geometric mean and median of $\langle \mu \rangle = 5.85$ and
$6.16$, respectively (and roughly the same for haloes with dwarf galaxy or
galaxy cluster masses). We therefore adopt $\mu=6$.

The reader may note a resemblance to the enclosed luminosity of the S\'ersic
model given in equations~(\ref{L3ofr}) and (\ref{L3tilde}), which may be
related to the resemblance of the projected NFW model to the S\'ersic profile
noted by \cite{LM01}, and the value of $\mu$ 
is of the rough order of the S\'ersic $m$'s (see
Fig.~\ref{figcor}).

Expressing the mean density at the virial radius $r_v = c\,r_{-2}$, one
obtains
\begin{equation}
{\rho_{-2}\over \rho_{\rm crit}} = {2\,\Delta\over3}\,
\left (2\,\mu\right)^{3\mu-1}\,\exp(-2\mu)\,{c^3\over\gamma\left (3\mu,2\mu
c^{1/\mu} \right )} \equiv f_1(c) \ .
\label{rhominusoverrhocritofmu}
\end{equation}
Now, \citeauthor{Navarro+04} provide a table with
$\rho_{-2} / \rho_{\rm crit}$ and $r_{-2}$.
We find that their data can be fitted by
\begin{equation}
{\rho_{-2}\over \rho_{\rm crit}}
\simeq \hbox{dex} 
\left [4.37 
- 0.30 \log \left ({h r_{-2} \over 1\, {\rm kpc}} \right )
- 0.10 \log^2 \left ({h r_{-2} \over 1\, {\rm kpc}} \right ) \right ] \equiv
f_2\left ({h r_{-2} \over 1\, {\rm kpc}} \right ) \ .
\label{rminus2fit}
\end{equation}
We solve this 2nd order polynomial for 
$h r_{-2} / (1\, {\rm kpc}) = f_2^{-1} (\rho_{-2}/\rho_{\rm crit})$.
Equations~(\ref{rhominusoverrhocritofmu}) and (\ref{rminus2fit}) can be
combined into an implicit equation for the concentration parameter $c$:
\begin{equation}
c\,f_2^{-1} \left [f_1(c) \right ] = r_v \ .
\label{cimplicit}
\end{equation}

Given that the virial radius is a function of the mass $M_v$ within the virial
radius (eq.~[\ref{rvir}]), then for a given $M_v$, one can solve
equation~(\ref{cimplicit}) for the
concentration parameter.
For $\mu=6$, we obtain
\begin{equation}
c \simeq 8.1\,M_{12}^{-0.11-0.015\,\log M_{12}} \,
\label{cNav04fit}
\end{equation}
where
$M_{12} = {h M_v  / 10^{12} {\rm M_\odot}}$.
\begin{figure}
\centering
\includegraphics[width=0.5\hsize]{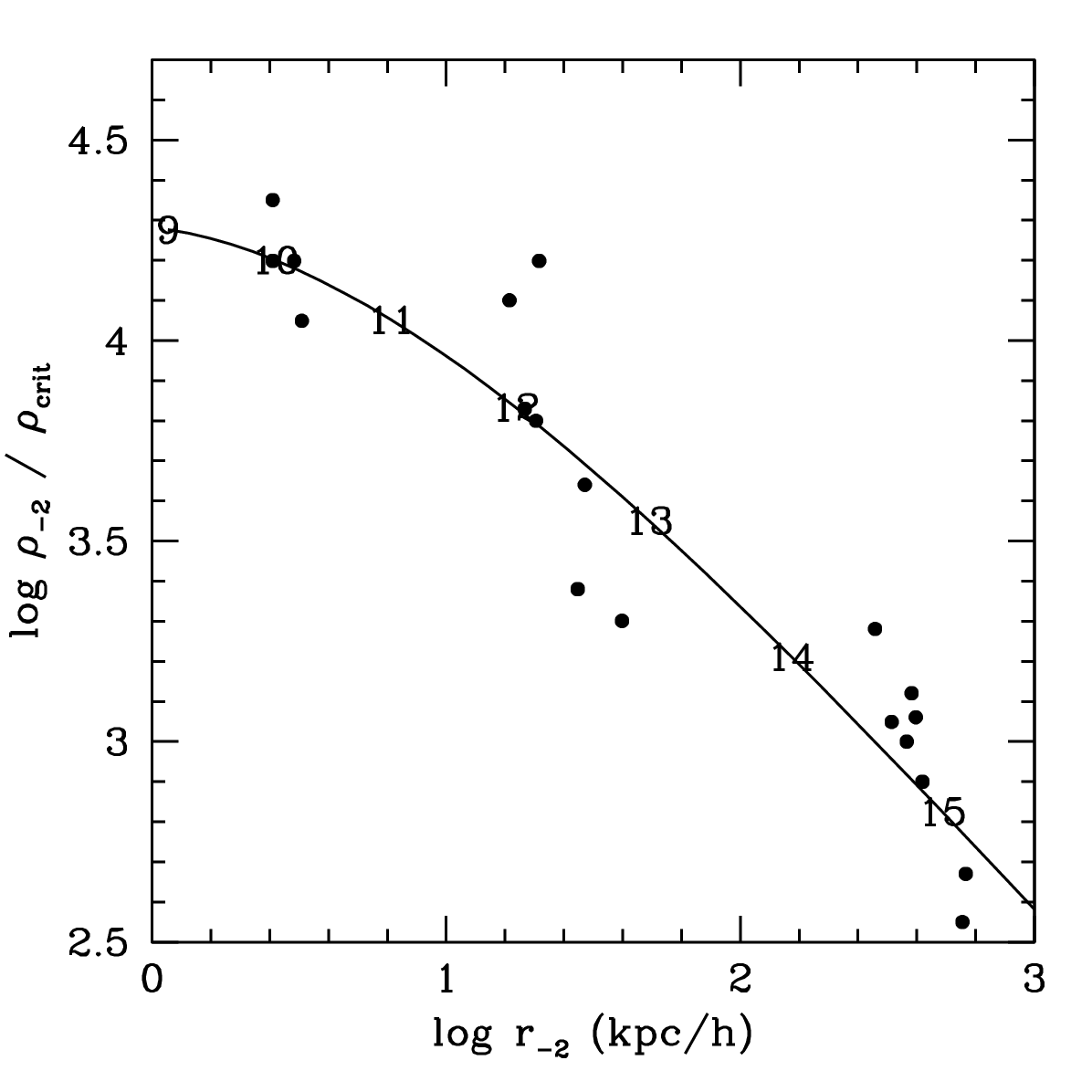}
\caption{Normalised density at scale radius versus scale radius for the 
Navarro et al. (2004) model.
\emph{Circles}: measurements by 
Navarro et al.;
\emph{curve}: predictions from our concentration parameter / mass
approximation 
(eq.~[\ref{cNav04fit}]) also using equations~(\ref{rvir}) and 
(\ref{rhominusoverrhocritofmu}); 
\emph{numbers}: $\log h_{70}\,M_v$.}
\label{chkcofm}
\end{figure}

Fig.~\ref{chkcofm} displays a check of equation~(\ref{cNav04fit}):
for every $M_v$, we obtain $c$ from equation~(\ref{cNav04fit}),
$\rho_{-2}/\rho_{\rm crit}$ from equation~(\ref{rhominusoverrhocritofmu}),
but also $r_v$ from equation~(\ref{rvir}) and $r_{-2} = r_v/c$.
The agreement is excellent and much better than if we had instead directly
obtained
$f_2^{-1}$ from a fit of
$h\,r_{-2}/(1\,\rm kpc)$ as a function of $\rho_{-2}/\rho_{\rm crit}$.

The central density of the Nav04 model is
\begin{equation}
\rho_0 \equiv \rho(0) = \rho_{-2}\,\exp(2\,\mu)
\label{rho0}
\end{equation}
Using equation~(\ref{rhominusoverrhocritofmu}), equation~(\ref{rho0}) can be
expressed in terms of 
the critical density
\begin{equation}
{\rho_0\over \rho_{\rm crit}} = {2\,\Delta\over
3}\,\left (2\,\mu \right)^{3\mu-1}\,{c^3\over\gamma\left (3\mu,2\mu\,
c^{1/\mu} \right )}
\end{equation}
and with equation~(\ref{cNav04fit}), this yields
\begin{equation}
\log \left ({\rho_0\over \rho_{\rm crit}} \right ) \simeq
9.0 - 0.25\,\log M_{12}-0.030\,\log^2 M_{12} \ .
\end{equation}
Thus the central density decreases with increasing mass.

The mass of the Nav04 model converges at infinity, and the mass at the virial
radius satisfies
\begin{equation}
{M_v \over M_\infty} = {\gamma(3\mu,2\mu\,c^{1/\mu}) \over \Gamma(3\mu)}
\simeq 0.44\,M_{12}^{-0.068-0.014\,\log M_{12}-0.0012\,\log^2 M_{12}} \ ,
\label{mfracvir}
\end{equation}
so that for giant galaxies a little over half the mass is beyond the virial
radius, while for rich clusters more than 3/4 of the mass is beyond $r_v$.

\section{Line-of-sight, aperture and slit 
velocity dispersions of isotropic systems}
\label{appJeans}
\subsection{Radial velocity dispersion}

The Jeans equation is
\begin{equation}
{{\rm d}(\ell \sigma_r^2) \over {\rm d}r} + 2\,{\beta(r) \over r} \,\ell\sigma_r^2 = -
\ell(r)\,
{GM(r)\over r^2} \ ,
\label{jeans1}
\end{equation}
where the anisotropy parameter is $\beta = 1 - \sigma_r^2/\sigma_t^2$, with
$\sigma_t = \sigma_\theta = \sigma_\phi$ the 1D tangential velocity dispersion,
so that $\beta = 0$ corresponds to isotropy, $\beta = 1$ is fully radial
anisotropy, and $\beta \to -\infty$ is fully tangential anisotropy.

For isotropic orbits, the Jeans equation~(\ref{jeans1}) trivially leads to
\begin{equation}
{\ell(r)\,\sigma_r^2 (r)\over G} = \int_r^\infty \ell \,M \,\left
({{\rm d}s\over s^2} \right ) \ .
\label{sigriso}
\end{equation}

\subsection{Line-of-sight velocity dispersion}

Projecting  along the line-of-sight, one trivially finds that
 the line-of-sight velocity
dispersion is
\begin{equation}
I(R)\,\sigma_{\rm los}^2 (R) = 2\,\int_R^\infty {\ell\sigma_r^2
\,r\,{\rm d}r\over \sqrt{r^2\!-\!R^2}} 
\ .
\label{siglos1}
\end{equation}
Inserting equation~(\ref{sigriso}) into equation~(\ref{siglos1}), and
inverting the order of integration, one easily finds (Prugniel \&
Simien 1997):
\begin{equation}
I(R) \,\sigma_{\rm los}^2 (R) = 2\,G\,\int_R^\infty 
{\sqrt{r^2\!-\!R^2}\over r^2}\ \ell(r) \,M(r)\,{\rm d}r \ .
\label{siglosisogen}
\end{equation}


Writing $M(r) = M_v\,\widetilde M(r/a_S)$,
$x = r/a_S$, $x_v = r_v/a_S$, 
$\eta = a_S / a_h$, 
the deprojected luminosity $\widetilde L_3$ of the S\'ersic profile is given in
equation~(\ref{L3tilde}),
the isotropic velocity dispersion, derived from
equation~(\ref{siglosisogen}), using
equations~(\ref{Isersic}),
(\ref{nuofr}), 
(\ref{nuofx}), 
(\ref{nu1}),
and (\ref{Mdarkofr})
becomes
\begin{equation}
\sigma_{\rm los}^2 (R) =  {\Gamma(2m)\over \Gamma[(3\!-\!p)m]} \,
{r_v \over a_S} \,V_v^2 \, \exp\left (X^{1/m} \right )
\int_X^\infty 
{\sqrt{x^2\!-\!X^2}\over x^2} \ 
\widetilde \ell(x) \,
\widetilde M(x)
\,{\rm d}x \ ,
\label{siglosiso}
\end{equation}
where$V_v^2 = G M_v/r_v$ is the squared circular
velocity at the virial radius and again $X = R / a_S$.

\subsection{Aperture velocity dispersion}


The aperture velocity dispersion satisfies
\begin{equation}
L_2(R)\,\sigma_{\rm ap}^2 (R) = \int_0^R 2\pi\,S\,I(S)\,\sigma_{\rm
los}^2(S)\,{\rm d}S \ ,
\label{sigapdef}
\end{equation}
where $L_2(R) = \int_0^R 2\pi S I(S)\, {\rm d}S$ is the luminosity within projected
radius $R$.
Inserting $I \sigma_{\rm
los}^2$ from equation~(\ref{siglosisogen}) into equation~(\ref{sigapdef}),
again inverting the order of integration,
we obtain the
isotropic solution for the aperture velocity dispersion\footnote{The number 3  (in
  red)
  in the denominator of the fraction before the square brackets in equation~(\ref{sigap1})  was erroneously omitted in the published version.}:
\begin{eqnarray}
\sigma_{\rm ap}^2 (R) &=& {4\pi\,G\over \textcolor{red}{3}\,L_2(R)}\,\left [\int_0^\infty r\,
\ell(r) M(r)\,{\rm d}r - \int_R^\infty  {(r^2\!-\!R^2)^{3/2}\over r^2}\,\ell(r)
M(r)\,{\rm d}r \right ] 
\label{sigap1}\\
&=&
{L\over L_2(R)}\,\sigma_{\rm tot}^2 - {4\pi \,G\over
3\,L_2(R)}\,\int_R^\infty {\left (r^2\!-\!R^2 \right)^{3/2}\over
r^2}\,\ell(r)\,M(r)\,{\rm d}r \ ,
\label{sigapisogen}
\end{eqnarray}
which, in the limit $R\to\infty$, converges to the (one-dimensional) 
isotropic velocity dispersion, averaged over the entire galaxy:
\begin{equation}
\sigma_{\rm tot}^2 = {4\pi \,G\over
3\,L}\,\int_0^\infty r\,\ell(r)\,M(r)\,{\rm d}r \ .
\end{equation}


Using the formula for the projected luminosity of the S\'ersic profile 
\citep{GC97,BJ98,LGM99}
\begin{equation}
L_2(R) =
2\pi\,m\,I_0\,a_S^2\,\gamma \left [2m,\left ({R\over a_S}\right
)^{1/m}\right ]  \ ,  \label{L2ofR}
\end{equation}
the isotropic aperture velocity dispersion at radius $R = X\,a_S$
(eq.~[\ref{sigapisogen}]), together with
equations~(\ref{nuofr}), 
(\ref{nuofx}), 
(\ref{nu1}),
(\ref{L3ofr}),
(\ref{Mdarkofr}),
and
(\ref{L2ofR})
yields
\begin{eqnarray}
\sigma_{\rm ap}^2 (R)  &=&
{\Gamma(2m) / \Gamma[(3\!-\!p)\,m] \over 3\,m\,\gamma \left (2m, X^{1/m} \right
)}\,{r_v\over a_S}\, V_v^2 
\left \{
\int_0^\infty \!\! x\,\widetilde \ell(x)
\,\widetilde M(x)\,
{\rm d}x
- \int_X^\infty {\left (x^2\!-\!X^2 \right )^{3/2}\over x^2}
\,\widetilde \ell(x)
\,\widetilde M(x)\,
{\rm d}x \right \} \ .
\label{sigapiso}
\end{eqnarray}

For the numerical integration of equation~(\ref{sigapiso}),
the first integral can be evaluated by integrating along $\ln x$ in the
range
$[-20,\ln x_{\rm max}]$, where $x_{\rm max} = 50^m$, for which the
exponential term in $\widetilde \ell (x)$ is extremely small.
The other integrals can be numerically evaluated by integrating $\ln (x/X)$
in
the range $[0,\ln (x_{\rm max} / X)]$.
In this paper, we evaluate numerically 
all velocity dispersions using {\tt Mathematica}.

\subsection{Slit-averaged velocity dispersion}

The velocity dispersion averaged over a thin slit of width $R$ is
\begin{equation}
\sigma_{\rm slit}^2 (R) = {\int_0^R I(S)\, \sigma_{\rm los}^2 (S)\,{\rm d}S \over
\int_0^R I(S)\, {\rm d}S } \ .
\label{sigslitdef}
\end{equation}
With $J(R) = \int_0^R I(S)\,{\rm d}S$,
equation~(\ref{sigslitdef}) becomes
\begin{eqnarray}
{J(R)\,\sigma_{\rm slit}^2 (R) \over G} &=& 
2\,\int_0^R {\rm d}S\,\int_S^\infty \ell\, M \,\sqrt{r^2\!-\!S^2}\,{{\rm d}r\over r^2}
\nonumber \\
&=& {\pi\over2}\,\int_0^R \ell(r)\, M(r)\,{\rm d}r + R\,\int_R^\infty \ell(r)\,
M(r)\,\sqrt{r^2\!-\!R^2}\,{{\rm d}r\over r^2} + \int_R^\infty \ell(r)\, M(r)\,
\sin^{-1} \left ({R \over r}\right)\,{\rm d}r \ ,
\label{sigslitisogen}
\end{eqnarray}
where 
the second equality in equation~(\ref{sigslitisogen}) was obtained after
inverting the order of integration in the first equality.
With equation~(\ref{Isersic}), 
one has
\begin{equation}
J(R) = m\,I_0\,a_S\,
\gamma \left [m, \left ({R\over a_S} \right )^{1/m} \right ] 
 .
\label{Jsersic}
\end{equation}
The isotropic  slit velocity dispersion (eq.~[\ref{sigslitisogen}])
becomes, with equations~(\ref{nuofr}), 
(\ref{nuofx}), 
(\ref{nu1}),
(\ref{Mdarkofr}),
and
(\ref{Jsersic}), 
\begin{eqnarray}
\sigma_{\rm slit}^2 (R) &=& {\Gamma(2m) / \Gamma[(3\!-\!p)\,m] \over
2\,m\,\gamma \left (m, X^{1/m} \right 
)}\,{r_v\over a_S}\, V_v^2 \nonumber \\
&\mbox{}& \times \left \{
{\pi\over2}\,
\int_0^X 
\widetilde \ell(x)\,
\widetilde M(x)\,
\,{\rm d}x
+ X\,
\int_X^\infty
{\sqrt{x^2\!-\!X^2}\over x^2}
\,\widetilde \ell(x)\,
\widetilde M(x)
\,{\rm d}x
+ \int_X^\infty 
\sin^{-1} \left ({X \over x}\right )
\ \widetilde \ell(x)\,
\widetilde M(x)
\,{\rm d}x
\right \}
\ ,
\label{sigslitiso}
\end{eqnarray}
where $X = R / a_S$.
The last two integrals are easily evaluated numerically with the substitution
$x = X/\sin u$.

\twocolumn

\bibliography{master}
\end{document}